\documentclass[smallextended]{svjour3}

\usepackage{algorithm}
\usepackage{algorithmicx}
\usepackage{algpseudocode}
\usepackage{amsmath}
\usepackage{amssymb}
\usepackage{bm}
\usepackage{graphicx}
\usepackage[numbers]{natbib}
\usepackage{balance}
\usepackage{xcolor}
\usepackage{booktabs}
\usepackage{wrapfig}
\usepackage{xspace}
\usepackage{natbib}
\usepackage{needspace}
\usepackage{subcaption}
\usepackage{listings}
\usepackage{tabularx}
\usepackage{booktabs}
\usepackage{colortbl}
\usepackage{multirow}
\usepackage{tcolorbox}
\tcbuselibrary{skins}
\usepackage{lmodern} 
\usepackage[absolute,overlay]{textpos} 
\usepackage[hyphens]{url}
\usepackage[colorlinks=false, breaklinks]{hyperref}
\usepackage[all]{nowidow}

\algnewcommand\algorithmicforeach{\textbf{for each}}
\algdef{S}[FOR]{ForEach}[1]{\algorithmicforeach\ #1\ \algorithmicdo}

\algnewcommand\algorithmicinput{\textbf{Input:}}
\algnewcommand\Input{\item[\algorithmicinput]}
\algnewcommand\algorithmicoutput{\textbf{Output:}}
\algnewcommand\Output{\item[\algorithmicoutput]}
\algrenewcommand\algorithmiccomment[1]{\State\textcolor{gray}{// #1}}

\algdef{SE}[REPEATN]{RepeatN}{EndRepeatN}[1]{\algorithmicrepeat\ #1 \textbf{times}}{\algorithmicend}

\newcommand*\card[1]{\lvert#1\rvert}
\DeclareMathOperator*{\deltatime}{\text{$\Delta t$}}

\newcolumntype{Y}{>{\centering\arraybackslash}X} 
\begin{document}
\begin{textblock*}{\textwidth}(2.5cm,23cm)
\footnotesize
This preprint has not undergone peer review (when applicable) or any post-submission improvements or corrections. The Version of Record of this article is published in Empirical Software Engineering (EMSE), and is available online at \url{http://dx.doi.org/10.1007/s10664-025-10737-8}.
\end{textblock*}

\title{Test Schedule Generation for Acceptance Testing of Mission-Critical Satellite Systems}
\titlerunning{Test Schedule Generation for Acceptance Testing of Mission-Critical Systems}

\author{Rapha{\"e}l Ollando \and Seung Yeob Shin \and Mario Minardi \and Nikolas Sidiropoulos}

\institute{
	Rapha{\"e}l Ollando \at
	{SnT, University of Luxembourg, Luxembourg} \\
	\email{raphael.ollando@uni.lu} \\
	\and
	Seung Yeob Shin \at
	{SnT, University of Luxembourg, Luxembourg} \\
	\email{seungyeob.shin@uni.lu} \\
	\and
	Mario Minardi \at 
	{SES Techcom, Luxembourg} \\
	\email{mario.minardi@ses.com} \\
	\and
	Nikolas Sidiropoulos \at
	{SES Techcom, Luxembourg} \\
	\email{nikolas.sidiropoulos@ses.com} \\
}

\date{Received: date / Accepted: date}

\maketitle

\begin{abstract}
Mission-critical system, such as satellite systems, healthcare systems, and nuclear power plant control systems, undergo rigorous testing to ensure they meet specific operational requirements throughout their operation.
This includes Operational Acceptance Testing (OAT), which aims to ensure that the system functions correctly under real-world operational conditions.
In satellite development, In-Orbit Testing (IOT) is a crucial OAT activity performed regularly and as needed after deployment in orbit to check the satellite's performance and ensure that operational requirements are met.
The scheduling of an IOT campaign, which executes multiple IOT procedures, is an important yet challenging problem, as it accounts for various factors, including satellite visibility, antenna usage costs, testing time periods, and operational constraints.
To address the IOT scheduling problem, we propose a multi-objective approach to generate near-optimal IOT schedules, accounting for operational costs, fragmentation (i.e., the splitting of tests), and resource efficiency, which align with practitioners' objectives for IOT scheduling.
Our industrial case study with SES Techcom shows significant improvements, as follows:
an average improvement of 49.4\% in the cost objective, 60.4\% in the fragmentation objective, and 30\% in the resource usage objective, compared to our baselines.
Additionally, our approach improves cost efficiency by 538\% and resource usage efficiency by 39.42\% compared to manually constructed schedules provided by practitioners, while requiring only 12.5\% of the time needed for manual IOT scheduling.
\end{abstract}

\keywords{Acceptance Testing, Mission-Critical Systems, Satellite Systems, Test Case Scheduling, Multi-Objective Optimization}

\section{Introduction}

Mission-Critical Systems (MCSs), such as satellite systems, healthcare systems, or nuclear power plant control systems, are developed and rigorously tested to ensure they meet specific operational requirements before being put into operation.
Furthermore, MCSs require additional testing phases during their lifespan, which are referred to as \emph{Operational Acceptance Testing} (OAT)~\cite{AmmannO2016}.
OAT is essential for MCSs to ensure that they meet all specified operational requirements under real-world conditions and are ready for sustained, reliable operation.

In satellite development and operation, \emph{In-Orbit Testing} (IOT) is an important OAT activity.
IOT is routinely performed following the successful deployment of a satellite, during which various subsystems of the satellite are tested in orbit.
The purpose of IOT is to compare the performance of the satellite with its pre-launch data and test results, ensuring that no degradations has occurred due to the stresses of launch or the environmental conditions in space~\cite{FortescueSS2011}.
Hence, IOT serves to confirm that the satellite's readiness to continue its mission.

IOT involves scheduling the testing campaign, including a test suite consisting of various test procedures to be performed on the satellite under test and determining when they should be performed.
This scheduling is inherently complex, as it must account for several factors, such as the frequency and duration of the satellite's visibility to a specific ground antenna, the costs associated with antenna usage, and the time required to configure and orient the antenna before each test.
In addition, scheduling IOT campaigns for a constellation of satellites presents additional challenges due to conflicts that may arise among test procedures for different satellites as a result of resource contention, temporal constraints, and other dependencies.
Furthermore, since satellite operators are typically responsible for both IOT campaigns and a variety of other satellite operation tasks, frequent transitions between IOT procedures and these operational responsibilities can significantly increase their cognitive load, thereby posing a risk of errors.
Hence, the complexity of scheduling IOT campaigns further increases to address the overhead caused by the fragmented execution of IOT procedures, which are interrupted by other operational tasks.

Regarding the IOT scheduling problem, the most relevant existing works are those on test case prioritization, which have been widely studied in software engineering~\cite{ArrietaWSE2016,WangAYOL2016,ShinNSBZ2018,ArrietaWSE2019}.
Among these, the most pertinent prior studies address the problem of test case prioritization in the context of Cyber-Physical Systems (CPSs).
For example, \citet{ShinNSBZ2018} proposed an automated test case prioritization method for CPSs that account for time budget constraints, uncertainties, and hardware damage risks.
In addition, other factors, such as fault detection time, simulation time, and requirement coverage, have also been considered in other prior studies~\cite{ArrietaWSE2016,WangAYOL2016,ArrietaWSE2019} on prioritizing test cases for CPSs. 
However, these existing techniques do not account for the specific characteristics of IOT for satellites, which include shared antenna usage, satellite visibility, operator involvement, and potential conflicts among IOT procedures.
Hence, they are not suitable for addressing the IOT scheduling problem.
In the satellite domain, although the satellite control resource scheduling problem (SCRSP) and ground measurement and control resource allocation (GMCRA) have been studied extensively in prior works~\cite{MarinelliNRS2011, ZhangZK2011, GaoWZ2013, WuLMQ2013, ZhangZF2014, ZhangHZ2018}, their methods primarily address problems related to satellite communication requests.
However, the problem of scheduling IOT campaigns has received relatively less attention.
Hence, in practice, IOT operators manually schedule IOT campaigns based on their expertise, which is time-consuming and prone to errors.

\textbf{Contributions}.
This article addresses the problem of scheduling IOT campaigns in an efficient and effective manner.
Specifically, our contributions are as follows:
(1)~\emph{A multi-objective approach to scheduling acceptance tests (i.e., IOT campaigns) for mission-critical satellite systems}.
Our approach includes
(a)~a precise definition of the problem of scheduling IOT campaigns, accounting for schedule objectives and constraints;
(b)~an algorithm based on Non-dominated Sorting Genetic Algorithm III (NSGA-III~\cite{DebJ2014:NSGA-III}) for finding near-optimal feasible IOT schedules; and
(c)~fitness functions that evaluate IOT schedules by assessing their operational cost, fragmentation (as fragmented IOT schedules incur overheads), and efficiency in the use of test resources.
(2)~\emph{An industrial case study}.
We applied our approach to a Global Navigation Satellite System (GNSS), for which SES Techcom, our industrial partner, provides operational services.
Our results show that an IOT campaign scheduled using our search-based approach, compared to a random search approach, finds feasible schedules that achieve an average improvement of 49.4\% in the cost fitness, 60.4\% in the fragmentation fitness, and 30\% in efficiency of the test resource usage fitness.
In addition, our approach demonstrates superior performance compared to an IOT scheduling approach based on Ant-Colony Optimization (ACO), which has been widely adopted in many prior studies addressing optimization and scheduling problems~\cite{BellM2004,KongTK2008,ZhangZK2011,GaoWZ2013,ZhangZF2014,ZhangHZ2018}.
Specifically, our approach outperforms an ACO-based approach by 53.1\% in cost efficiency, 58.3\% in fragmentation, and 26.1\% in efficiency of resource usage over the same period.
Moreover, our approach yields schedules that improve the cost efficiency by 538\%, and the efficiency of the test resource usage by 39.42\% compared to schedules manually constructed by practitioners, while maintaining comparable performance in terms of fragmentation and requiring only 12.5\% of the time needed by practitioners to construct an IOT schedule.
(3)~\emph{Practitioners' feedback on our IOT scheduling approach}.
Finally, we gathered feedback on our approach from practitioners at SES Techcom during the delivery of our research outcomes.
They highlighted the following:
(a)~the efficiency of schedule generation, as our automated approach generates feasible schedules much faster than manual methods, enabling quick adaptation to changing conditions and needs, and
(b)~the ability to produce several equally viable schedules, facilitating trade-off analysis.

\textbf{Organization}.
In Section~\ref{sec:background}, we provide the background of this research.
Section~\ref{sec:approach} describes our approach to generate schedules for acceptance testing of mission-critical satellite systems.
In Section~\ref{sec:evaluation}, we perform our empirical evaluation, discuss threats to validity, and present lessons learned from practitioners' feedback.
Finally, Section~\ref{sec:conclusions} concludes the article.
 \section{Background}
\label{sec:background}

\subsection{Motivating Case Study}

We motivate our work using a case study from SES Techcom, which develops satellite-enabled solutions.
Operators of satellites are tasked with ensuring optimal performance of their satellites' services once deployed in orbit.
Given the critical role of satellite technology in supporting various services, such as broadcast television, global navigation and positioning systems, and mobile communications, operators must ensure that, over the lifespan of a satellite, the Quality-of-Service (QoS) remains within the standards defined by its application.
Consequently, operators routinely conduct \emph{In-Orbit Testing} (IOT) procedures to monitor the QoS of each satellite in the constellation they operate.
These IOT procedures have four main objectives:
(1)~ensuring the behavior of the satellite remains consistent before and after launch,
(2)~verifying performance adherence to specifications, 
(3)~forecasting end-of-life,
and (4)~investigating potential anomalies.

SES Techcom conducts routine monthly tests for the European GNSS constellation, Galileo.
In this context, \emph{IOT} procedures are divided into two categories: \emph{Signal Quality Monitoring} (SQM) and \emph{Routine In-Orbit Test} (RIOT).
SQM procedures measure the satellite's signal quality and strength on each communication channel.
Specifically, these procedures involve measuring the Modulated Effective Isotropic Irradiated Power (EIRP)~\cite{Balanis2016, MaralBS2020} approximately 15 times on each channel, with the overall testing duration lasting almost one hour per satellite.
The SQM procedures are usually performed at the highest elevation available at any given pass of a satellite. 
RIOT procedures are performed sequentially throughout the full pass of the satellite, from the signal acquisition, typically around 3-5 degrees of elevation, until signal loss at a similar elevation.
The duration of an RIOT phase ranges from 8 to 9 hours, depending on the satellite and the ground measuring station.
During an RIOT phase, several IOT measurements are performed for every Galileo channel.
Specifically, these IOT measurements include Modulated EIRP, IQ sample collection, out-of-band spurious measurement, navigation receiver data analysis, and Search-and-Rescue (SAR) check, if SAR is available~\cite{MaralBS2020}. 

Additionally, the antennas used to communicate with the satellite are large objects that require time to be precisely pointed toward the satellite under test.
Due to the precise nature of satellite communication, test instruments and antenna alignment often need to be re-calibrated before conducting each test procedure.
These factors introduce delays before conducting each test procedure in an IOT campaign, during which no tests can be performed, and must therefore be taken into consideration in the scheduling process.

Currently, practitioners at SES Techcom manually schedule these \emph{IOT} procedures, having determined that existing automated solutions are not practically applicable to their scheduling needs.
However, this manual approach poses significant challenges and consumes valuable time for practitioners, particularly when the satellites' orbits have short revolution periods or substantial inclinations.
Moreover, in the event of an emergency scenario, such as an unexpected degradation in QoS across the constellation, an IOT campaign must be scheduled and executed within a condensed time-frame.
With the Galileo constellation currently consisting of 30 satellites in orbit (soon to be 32), this presents a considerable challenge for the IOT operators.
Hence, an algorithm that automatically solves the problem of scheduling IOT campaigns in practical time is highly desirable.

\subsection{IOT Requirements and Constraints}
\label{sec:requirements and constraints}

To create a suitable schedule for an IOT campaign, several key factors specific to the problem must be considered.

\noindent\textbf{Context switching.}
Minimizing context switching for satellite operators is essential for maintaining the efficiency and accuracy of the IOT campaign.
Frequent transitions between IOT procedures and other satellite operation tasks can increase cognitive load for operators, raising the likelihood of errors.
Additionally, context switching incurs time and resource costs, as operators must reorient themselves with each new task.
Streamlining workflows and grouping similar tasks can reduce the need for context switching.

\noindent\textbf{Utilization of test resources.}
Efficient use of IOT resources is crucial, ensuring that the equipment used for the IOT procedures (e.g., antennas, satellites, and test devices) is optimally utilized with minimal interruptions. 
Efficient IOT schedules ensure that these hardware resources are not overused, reducing not only operational costs but also the risk of hardware failures.
Inefficient test schedules increase the likelihood of hardware malfunctions due to several factors, such as overheating and exposure to harsh environmental conditions.
Additionally, efficient IOT schedules limit exposure to external disruptions, such as power outages, ensuring the integrity of the tests.
Studies show that the probability of hardware failure rises with continuous operation~\cite{OConnorK2012}.
By keeping the test campaign efficient, engineers can maintain optimal equipment performance and achieve more reliable results.

\noindent\textbf{Operational costs.}
The operational costs associated with performing IOT campaigns are significant and multifaceted.
These costs include the expenses related to the use of the antennas.
Additionally, there are costs associated with allocating human resources, including the personnel required to operate the IOT campaign.
Efficient management of these resources is essential for cost optimization. \section{Approach}
\label{sec:approach}

This section describes our approach to addressing the following problem:
For an IOT campaign to test satellites in a constellation, how can we create suitable IOT schedules that
(1)~enable efficient use of the antenna resources required for the test procedures in the IOT campaign,
(2)~reduce the frequency of context switching for operators conducting the IOT campaign, and
(3)~minimize the costs directly associated with executing the test procedures.

\subsection{IOT Scheduling Concepts}
\label{sec:scheduling-concepts}

To develop our approach, we define four concepts to find the most suitable schedule for an IOT campaign: \emph{satellite passes}, \emph{test procedures}, \emph{procedure schedules}, and \emph{slot schedules}.
Below, we precisely describe these concepts.

\paragraph{\textbf{Satellite pass.}}
Any satellite orbiting the Earth, except for those in geostationary orbit, can only be observed from a specific location on Earth during the period when the satellite passes in the visibility range of the ground station.
We refer to this period as a \emph{satellite pass}.
A satellite pass begins when the satellite rises above the horizon, reaches its zenith (highest elevation in the sky), and ends when it descends below the horizon.
During a pass, various activities such as communication, data collection, or observation activities between the satellite and ground stations can take place.
Furthermore, we define a pass of a satellite $s$ over a location $r$, denoted $\alpha_r^s$, as follows:
\begin{equation*}
    \alpha_r^s = \{ t_{start}, t_{max}, t_{end}, \theta_{start}, \theta_{max}, \theta_{end}, \phi_{start}, \phi_{max}, \phi_{end}\}
\end{equation*}
where $t_{start}$, $t_{max}$, and $t_{end}$ represent the time at which satellite $s$ begins its pass, reaches its maximum elevation, and finishes its pass at location $r$, respectively; $\theta_{start}$, $ \theta_{max}$, and $\theta_{end}$ represent the elevation angles at which $s$ begins its pass, reaches its maximum elevation, and finishes its pass at $r$ respectively; and $\phi_{start}$, $\phi_{max}$, and $\phi_{end}$ represent the azimuth angles at which $s$ begins its pass, reaches its maximum elevation and finishes its pass at $r$, respectively.
Similarly, we define $\Gamma_r^s(t_1, t_2) = \{\alpha_r^s \mid \alpha_r^s \text{ occurs between } t_1 \text{ and } t_2\}$ the set of satellite passes of satellite $s$ over location $r$ during a time period ranging from $t_1$ and $t_2$.
 
\paragraph{\textbf{Test procedure.}}
A \emph{test procedure} refers to a specific IOT procedure that is to be performed on a given satellite $s$.
A test procedure is characterized by a period during which the IOT procedure is conducted.
Formally, we define a test procedure associated with a satellite $s$, denoted $\tau_s$, as follows:
\begin{equation*}
    \tau_s = \{ t_{start}^s, t_{end}^s, \mathrm{Type}, \delta_c, \alpha_r^s\}
\end{equation*}
where $t_{start}^s$ and $t_{end}^s$ represent the start and end times of the test procedure, respectively,
$\mathrm{Type}$ represents the type of test procedure that is performed,
$\delta_c$ represents the configuration time required before performing the test procedure (e.g., repositioning the antenna, booting the equipment, etc.),
and $\alpha_r^s$ is the associated satellite pass.
We note that $t_{start}^s < t_{end}^s$, $t_{start} \leq t_{start}^s$, and $t_{end}^s \leq t_{end}$, where $t_{start} \in \alpha_r^s \in \tau_s, t_{start}^s \in \tau_s, t_{end}^s \in \tau_s,$ and $t_{end} \in \alpha_r^s \in \tau_s$.

In addition, we define the \emph{span} between two individual test procedures $\tau_i$ and $\tau_j$, where $\tau_i$ occurs before $\tau_j$, denoted $\Call{span}{\tau_i, \tau_j}$ as the elapsed time between the beginning of $\tau_i$ and the end of $\tau_j$, formally defined as follows:
\begin{equation*}
    \Call{span}{\tau_i, \tau_j} = \deltatime(t_{start}^i, t_{end}^j)
\end{equation*}

\paragraph{\textbf{Procedure schedule.}}
A \emph{procedure schedule} consists of a collection of test procedures over a defined time-frame. 
Formally, we define a procedure schedule, denoted $\mathcal{S}$, as follows:
\begin{equation*}
    \mathcal{S} = \{ \tau_1, \tau_2, \dots, \tau_n \}
\end{equation*}
where each $\tau \in \mathcal{S}$ corresponds to an individual test procedure, as defined previously.
Notably, we can define the \emph{span} of a schedule $\mathcal{S}$ as the time elapsed between the beginning of the first test procedure, and the end of the last test procedure in $\mathcal{S}$, denoted $\Call{span}{\mathcal{S}}$, and defined as follows:
\begin{equation*}
    \Call{span}{\mathcal{S}} = \deltatime\left(\min_{\tau_i \in \mathcal{S}}t_{start}^i, \max_{\tau_i \in \mathcal{S}}t_{end}^i\right)
\end{equation*}

\paragraph{\textbf{Slot schedule.}}
A \emph{slot schedule} refers to a collection of time slots during which an operator's resources (e.g., antennas and other equipment) are allocated for performing the IOT procedures defined in a procedure schedule.
Specifically, each procedure schedule $\mathcal{S}$ is associated with a unique slot schedule, denoted $\mathcal{Q}$, so that $\mathcal{Q} = \{(t_{j,start}, t_{j,end}) \mid j \in \{1, 2, \dots, n\}\}$, where $n$ is the number of time intervals, while $t_{j,start}$ and $t_{j,end}$ are the start and end time of the $j$-th interval, respectively.
\begin{figure}
    \centering
    \includegraphics[width=\linewidth]{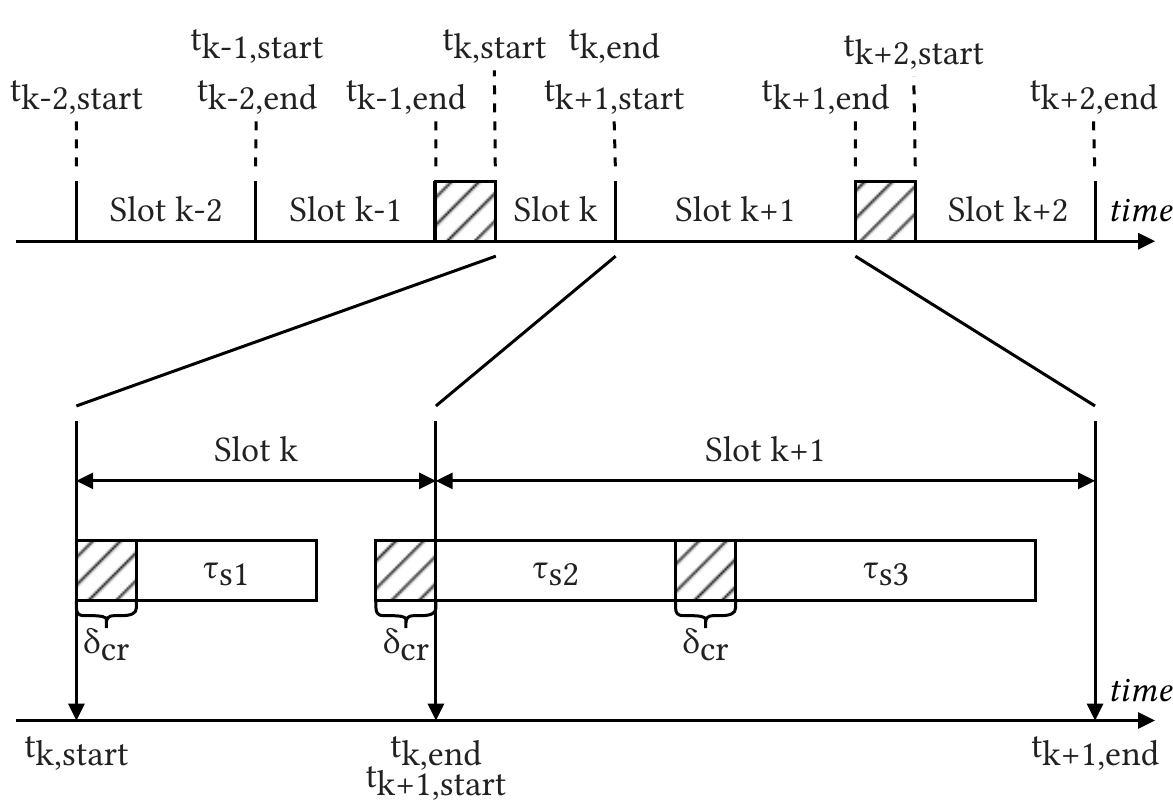}
    \caption{Example of a slot schedule and the corresponding slots.}
    \label{fig:slotting-example}
\end{figure}
Figure~\ref{fig:slotting-example} illustrates the relationship between a slot schedule and a procedure schedule.
The top portion of the figure depicts a slot schedule consisting of five slots, defined as $\mathcal{Q} = \{(t_{k-2,\text{start}}, t_{k-2,\text{end}}),$ $(t_{k-1,\text{start}}, t_{k-1,\text{end}}),$ $(t_{k,\text{start}}, t_{k,\text{end}}),$ $(t_{k+1,\text{start}}, t_{k+1,\text{end}}),\allowbreak (t_{k+2,\text{start}}, t_{k+2,\text{end}})\}$.
The bottom portion of the figure represents three test procedures, $\tau_{s1}$, $\tau_{s2}$, and $\tau_{s3}$, extracted from the procedure schedule $\mathcal{S}$.
This figure demonstrates that the slots encompass the procedures in $\mathcal{S}$.
Specifically, $\tau_{s1}$ is contained within the $k$-th slot of $\mathcal{Q}$, while $\tau_{s2}$ and $\tau_{s3}$ are contained within the $(k+1)$-th slot. 
Additionally, the figure highlights that slots do not need to be contiguous or temporally aligned with test procedures.

\paragraph{\textbf{IOT schedule.}}
An \emph{IOT schedule} is the outcome of the scheduling process, encompassing both the procedure schedule and the slot schedule. Formally, an IOT schedule, denoted as $\mathcal{P}$ is defined as:
\begin{equation*}
\mathcal{P} = (\mathcal{S}, \mathcal{Q})
\end{equation*}
where $\mathcal{S}$ represents the procedure schedule and $\mathcal{Q}$ represents the slot schedule.

\subsection{Identifying Conflicting Test Procedures}
\label{sec:conflicts}

\subsubsection{Conflict definition}
As explained in Section~\ref{sec:background}, scheduling IOT procedures for a constellation of satellites is a challenging activity that involves several constraints.
When a test procedure is selected for scheduling, conflict may arise with other test procedures if they cannot be executed simultaneously due to resource contention, temporal constraints, or other dependencies.

As an example, consider an IOT campaign of the Galileo constellation, our test subject provided by SES Techcom, where a minimum of four satellites are constantly visible in the sky at all times.
The IOT campaign is subject to the following constraints: 
Only a single antenna is available for use, meaning that only one test procedure can be conducted at any given time.
The reconfiguration overhead, which includes the time required to program and orient the antenna before initiating a test procedure, is 15 minutes. 
SQM test procedures must be conducted for 45 minutes, centered around the satellite's highest elevation point.
Recall from Section~\ref{sec:background} that the purpose of an SQM is to detect potential hazardous deformations in the signal emitted by the satellite. 
Additionally, RIOT procedures are required to be performed for the entire duration of a satellite pass.
This duration is defined as the period when the satellite's elevation is between 5º at the start and end of the pass.
Recall from Section~\ref{sec:background} that, the purpose of an RIOT is to test various capabilities of the satellite.
These constraints imply that RIOT test procedures cannot be scheduled for concurrent testing with any other test procedures, and SQM test procedures may not be scheduled for concurrent testing either. 
This highlights why identifying conflicts is necessary for creating IOT schedules.

Let $\mathcal{T}$ be a set of test procedures involved in the creation of an IOT schedule.
The set $\mathcal{T}$ is constructed by engineers who assign to each satellite pass $\alpha^r_s \in \Gamma^s_r$ an SQM or a RIOT test procedure, if applicable.
We define, for each $\tau_i, \tau_j \in \mathcal{T}, \tau_i \neq \tau_j$, the ``conflict'' $\xi_{\tau_i, \tau_j}$ between $\tau_i$ and $\tau_j$ such that $\xi_{\tau_i, \tau_j} = 1$ when there is a conflict and $\xi_{\tau_i, \tau_j} = 0$, otherwise.
Subsequently, for a set of test procedures $\mathcal{T}$, we can define a set $\Xi(\mathcal{T})$ representing the tuples of test procedures that conflict with each other within $\mathcal{T}$, as follows:
\begin{equation*}
    \Xi(\mathcal{T}) = \{(\tau_i, \tau_j)  \mid  i, j \in \{1, 2, \dots, \card{\mathcal{T}}\}, i \neq j, \xi_{\tau_i, \tau_j} = 1 \}
\end{equation*}

\begin{figure}[t]
    \centering
    \includegraphics[width=\linewidth]{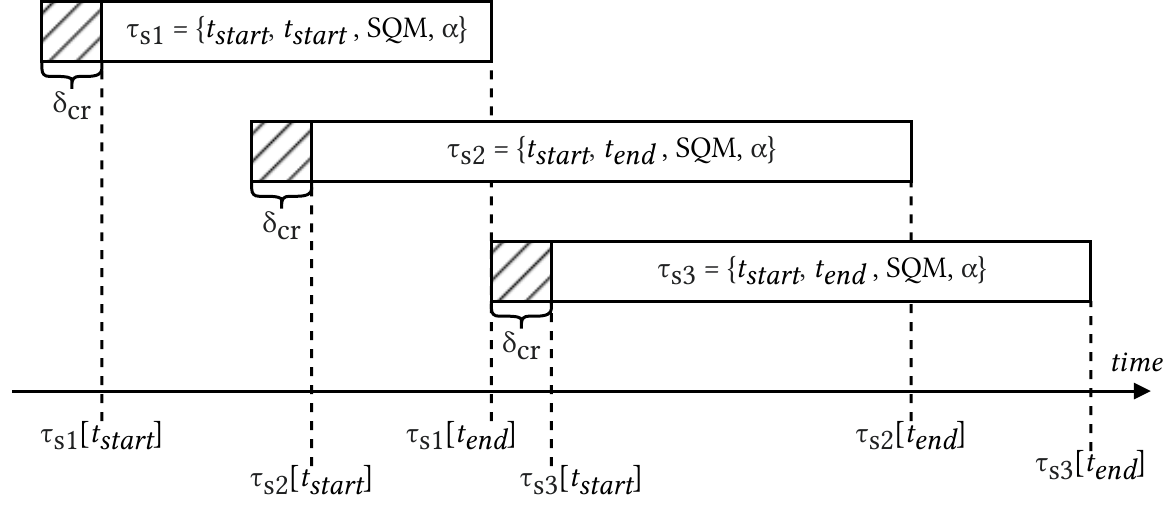}
    \caption{Three conflicting test procedures.}
    \label{fig:conflicting-tests}
\end{figure}
As an illustration, let us consider three test procedures $\tau_{s1}$, $\tau_{s2}$, and $\tau_{s3}$, depicted in Figure~\ref{fig:conflicting-tests}.
These test procedures have to be scheduled for three satellites $s_1$, $s_2$, and $s_3$, respectively, all from the same location $r$, and are of the identical type SQM.
We can see that the test procedure ${\tau_{s2}}$ starts after ${\tau_{s1}}$ begins but before ${\tau_{s1}}$ ends.
Consequently, there is a conflict between ${\tau_{s1}}$ and ${\tau_{s2}}$, resulting in $\xi_{\tau_{s1}, \tau_{s2}} = 1$.
Similarily, we observe a similar conflict between ${\tau_{s2}}$ and ${\tau_{s3}}$, resulting  in $\xi_{\tau_{s2}, \tau_{s3}} = 1$.
However, we can observe that no conflict occurs between ${\tau_{s1}}$ and ${\tau_{s3}}$, resulting in $\xi_{\tau_{s2}, \tau_{s3}} = 0$.
Thus, we have, for this set of test procedures, the following conflict set $\Xi(\mathcal{T}) = \{(\tau_1, \tau_2), (\tau_2, \tau_3)\}$.

\subsubsection{Conflict graph}
\label{sec:conflict-graph}

Based on the previous definition of conflicting test procedures, we elect to represent the conflicts in a set of test procedures as a \emph{conflict graph}.
Conflict graphs are undirected graphs $G = (V, E)$, where each vertex corresponds to a unique test procedure and each edge $(i, j) \in E$ represents the presence or absence of a conflict between a test procedure $i$ and a test procedure $j$~\cite{West2001:GraphTheory}.
Formally, for a set of test procedures $\mathcal{T} = \{ \tau_1, \tau_2, \dots, \tau_n \}$, we define a conflict graph as $G(\mathcal{T}) = (\mathcal{T}, \Xi({\mathcal{T}}))$, where $\mathcal{T}$ is the set of vertices of $G$ and $\Xi({\mathcal{T}})$ is the set of edges.
\begin{figure*}[t]
    \centering
    \includegraphics[width=\linewidth]{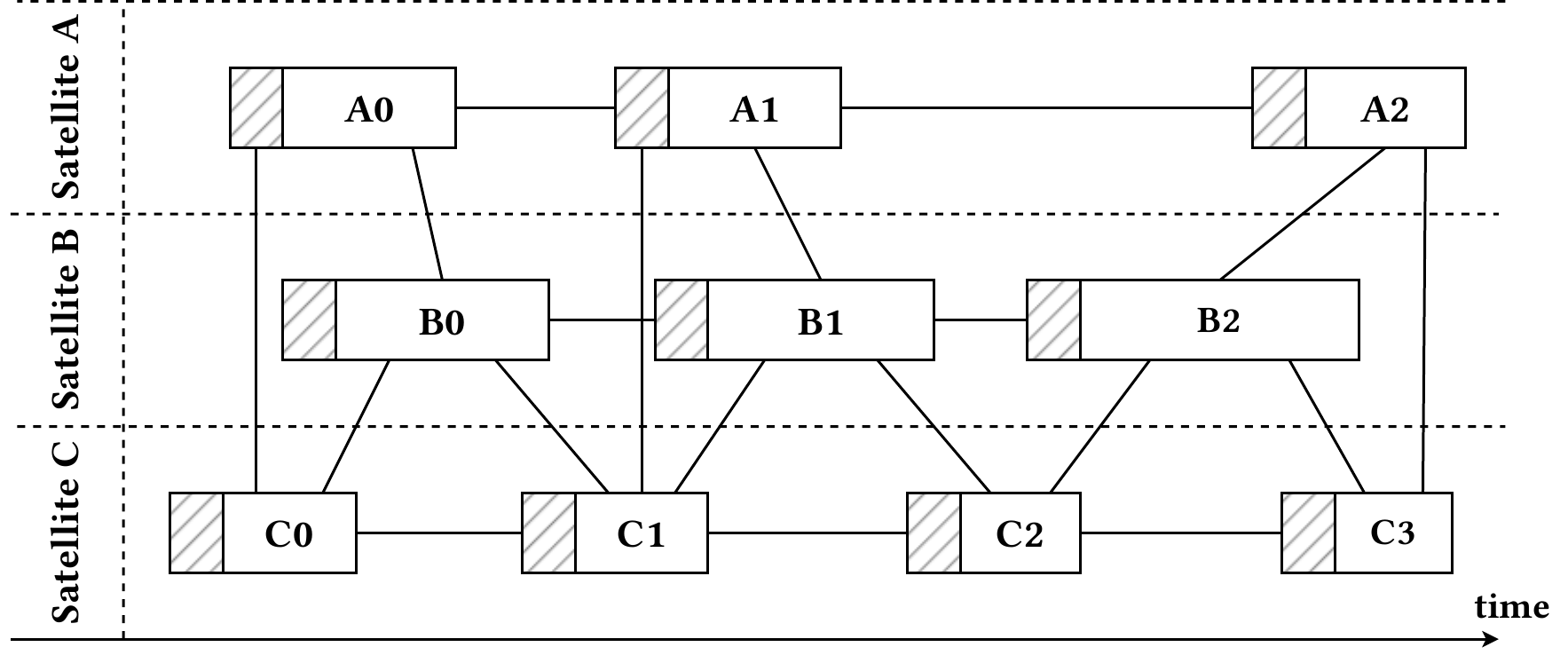}
    \caption{Example of a conflict graph created from the passes of three satellites (A, B, and C)}
    \label{fig:conflict_graph}
\end{figure*}
Figure~\ref{fig:conflict_graph} presents a conflict graph derived from the satellite passes of three satellites A, B, and C.
Each row of Figure~\ref{fig:conflict_graph} corresponds to candidate test procedures to be conducted for a given satellite.
For instance, in the first row, $\tau_{A0}$, $\tau_{A1}$, and $\tau_{A2}$ represent three separate candidate test procedures for satellite A.
The test procedures are organized horizontally according to their time of occurrence.
For instance, $\tau_{A0}$ occurs before $\tau_{A1}$, but $\tau_{A0}$ overlaps with $\tau_{B0}$ and $\tau_{C0}$.
Each hatched square represents the required configuration time $\delta_{cr}$ for each task in the graph.
Edges between nodes indicate conflicts between pairs of test procedures, meaning those procedures cannot be part of the same IOT schedule.
For example, an edge exists between $\tau_{A0}$, $\tau_{A1}$, and $\tau_{A2}$ because they are different candidate test procedures of the same type for satellite A, and thus cannot be part of the same candidate IOT schedule.
Similarly, an edge exists between $\tau_{A1}$ and $\tau_{B1}$ because these test procedures overlap in time, preventing their simultaneous execution.
An edge also exists between $\tau_{B2}$ and $\tau_{C2}$ because, although they do not overlap in time, the configuration time required for $\tau_{B2}$ does not allow sufficient time for the antenna to be repositioned after $\tau_{C2}$.
Using such a graph representation allows us to assess the feasibility of a procedure schedule efficiently.
To know if a schedule is feasible, we simply use the relation
\begin{equation*}
    \mathit{feasible}(\mathcal{S}) =
    \begin{cases}
        1 & \nexists \tau_i,\tau_j \in \mathcal{S}, (\tau_i,\tau_j) \in \Xi(\mathcal{T}) \\
        0 & \text{Otherwise}
    \end{cases}
\end{equation*}
For example, in the graph depicted in Figure~\ref{fig:conflict_graph}, a feasible procedure schedule would be $\mathcal{S} = \{\tau_{A1}, \tau_{B0}, \tau_{C2}\}$ as none of those test procedures possesses an edge connecting it to another test procedure in the graph.

\subsection{Schedule Optimization}
\label{sec:schedule optimization}

In this section, we present our approach to optimizing the scheduling for an IOT campaign.
Let $t_{sc}$ and $t_{se}$ be the start time and end time of the IOT campaign,
$r$ be the site on which the campaign is performed,
$S_{sat} = \{s_1, s_2, \dots, s_n\}$ be the set of the satellites to perform the IOT campaign on,
$\Gamma = \{\Gamma_r^s(t_{sc}, t_{se}) \mid s \in S_{sat}\}$ be the set of all satellite passes that will occur over $r$ for each satellite of $S_{sat}$ during the IOT campaign,
and $\mathcal{T} = \{\tau_1, \tau_2, \dots, \tau_n\}$ be the set of all possible test procedures that can be scheduled for the satellite passes in $\Gamma$.

We aim at finding a complete procedure schedule $\mathcal{S}$, i.e., a set of test procedures, that covers all the satellites and IOT procedures types to be performed, such that they are the (near-)optimal given the objectives described in the introduction of this section: the procedure schedules should (1)~maximize the antenna resource usage, (2)~reduce the number of context switching required from practitioners, and (3)~minimize the monetary and organizational cost of executing such schedules.
Once engineers obtain a set of equally viable and near-optimal procedure schedules according to the objective described above, they can select a single schedule for the IOT campaign.
This selection is made at the engineer's discretion, considering internal constraints, such as the availability of an IOT operator.
Note that in our approach, a slot schedule is determined by a procedure schedule.
We cast our problem of finding such procedure schedules into a \emph{multi-objective search optimization problem}~\cite{Luke2013:Metaheuristics}.
Following common practices for expressing multi-objective search problems, we define the representation of a (candidate) solution, the fitness functions used for evaluation, and the computational search algorithm employed to find near-optimal solutions.

\subsubsection{Representation}
Given a set $\mathcal{T}$ of possible test procedures, a solution of the optimization problem represents a subset $\mathcal{S} = \{\tau_1, \tau_2, \dots, \tau_n\}$, where $\tau_i \in \mathcal{S}$ and $\tau_i \in \mathcal{T}$.
To ensure that each procedure is unique and all combinations of pairs of type and satellite $\{\text{Type}, s\}$ specified by the problem are covered by distinct test procedures in $\mathcal{S}$, the following conditions must be met for all distinct $\tau_i, \tau_j \in \mathcal{S}$: (1)~the type, start time, and end time of $\tau_i$ are different from those of $\tau_j$,
and (2)~for every combination $\{\text{Type}, s\}$ specified by the problem, there exists a unique $\tau_k \in \mathcal{S}$ such that $\text{Type}_k = \text{Type}$ and $\alpha_r^k$ is a satellite pass of $s$.

\subsubsection{Fitness functions}
\label{sec:fitness_functions}

Our approach aims at searching for candidate procedure schedules with regard to three objectives: ($\mathcal{O}_1$)~maximizing the resource usage efficiency, ($\mathcal{O}_2$)~minimizing the context switching required from practitioners, and ($\mathcal{O}_3$)~minimizing the monetary and organizational cost of executing such a schedule.
To quantify how a candidate solution fits these three objectives, we define the following fitness functions:

\noindent\textit{Resource usage efficiency ($\mathcal{O}_1$).}
Recall from Section~\ref{sec:background} that engineers aim at efficiently using IOT resources, particularly antennas.
In most instances, engineers prioritize IOT schedules that are both short in duration and maximize the antenna usage over that time.
Note that if test procedures in a schedule are conducted with minimal idle time, antenna usage during IOT time is maximized, and the schedule requires the minimum possible time.
Thus, maximizing the usage of the antenna over that duration results in an efficient procedure schedule, minimizing the time required to complete an IOT campaign.
We define an antenna efficiency metric.
This metric estimates how much the antenna is used over the complete duration of a procedure schedule, and should be maximized.

Let $\mathcal{S}$ be a candidate schedule, $\delta_{rc}$ the reconfiguration time required between two consecutive test procedures, $t_{start}$ and $t_{end}$ be the start time and end time of the test procedure $\tau$, and $\deltatime(\tau)$ be the duration of a test procedure $\tau$.
Based on these definitions, we define the fitness function for objective $\mathcal{O}_1$, denoted $fituse(\mathcal{S})$, as follows:
\begin{equation*}
    \mathit{fituse}(\mathcal{S}) = \frac{1}{\Call{span}{\mathcal{S}}}  \left( \left( \card{\mathcal{S}}-1 \right) \delta_c + \sum\limits_{\tau \in \mathcal{S}}\deltatime(\tau) \right)
\end{equation*}
$\mathit{fituse}(\mathcal{S})$ is calculated as the inverse of the total schedule span, $\Call{span}{\mathcal{S}}$, multiplied by the sum of reconfiguration times between consecutive test procedures and the total duration of all test procedures in the schedule. 
The maximum value of $\mathit{fituse}(\mathcal{S})$ ($=1$)  is achieved when all test procedures $\tau \in \mathcal{S}$ are scheduled consecutively without any idle time.
Inversely, the minimum value occurs if the idle time between test procedures becomes infinitely large.

For example, let us consider a procedure schedule $\mathcal{S} = \{\tau_1, \tau_2, \tau_3\}$.
If $\Call{Span}{\mathcal{S}} = 15h$ , with each reconfiguration time $\delta_{rc} = 1h$, and the durations of the test procedures being $\deltatime(\tau_1) = 3h$, $\deltatime(\tau_1) = 4h$, and $\deltatime(\tau_1) = 2h$ respectively, then:
\begin{equation*}
    \mathit{fituse}(\mathcal{S}) = \frac{1}{15} \left( (3-1) \times 1 + (3 + 4 + 2) \right) = \frac{1}{15} \times 12 = 0.8
\end{equation*}

\noindent\textit{Minimizing context switching ($\mathcal{O}_2$).}
As explained in Section~\ref{sec:background}, in an IOT campaign, having too many time slots results in various impacts due to switching overhead.
On the contrary, adopting longer, consolidated periods in a slot schedule can enhance resource utilization by minimizing setup and teardown activities, reducing idle time, and maximizing the utilization of equipment, facilities, and operators.
Hence, our second fitness function $\mathcal{O}_2$, denoted as $\mathit{fitfrag}(\mathcal{S})$, ensures that the IOT schedule possesses a slot schedule that involves as few context switching as possible.

We note that, in our approach, a slot schedule $\mathcal{Q}$ is determined based on a procedure schedule $\mathcal{S}$.
However, creating a slot schedule depends on the operational context of each company conducting IOT. 
Hence, in Section~\ref{sec:evaluation}, we present the slot scheduling algorithm, which takes as input $\mathcal{S}$ and outputs $\mathcal{Q}$, specific to the context of our experiments.

Given a candidate procedure schedule $\mathcal{S}$, we define the fitness function $\mathit{fitfrag}(\mathcal{S})$, as follows:
\begin{equation*}
    \mathit{fitfrag}(\mathcal{S}) = 1 - \frac{\card{\mathcal{Q}} - 1}{\card{\mathcal{S}} - 1}
\end{equation*}
where $\card{\mathcal{Q}}$ is the number of separate time slots in the slot schedule $\mathcal{Q}$ for $\mathcal{S}$.
$\mathit{fitfrag}(\mathcal{S})$ reaches its maximum value ($=1$) when there is no fragmentation in $\mathcal{Q}$, i.e., all test procedures are scheduled consecutively under the same slot.
The function reaches its minimum value ($= 0$) when a unique slot is assigned to each test procedure individually.

For example, let us consider a procedure schedule $\mathcal{S}$ with four test procedures $\tau_1$, $\tau_2$, $\tau_3$, and $\tau_4$.
If there are two slots in the slot schedule, each covering two test procedures separately, then $\card{\mathcal{Q}} = 2$ and $\card{\mathcal{S}} = 4$.
The fitness is calculated as follows:
\begin{equation*}
    \mathit{fitfrag}(\mathcal{S}) = 1 - \frac{2 - 1}{4 - 1} = 1 - \frac{1}{3} = 0.667
\end{equation*}

\noindent\textbf{Minimizing the monetary and organizational cost ($\mathcal{O}_3$).}
Recall from Section~\ref{sec:background} that, in the context of IOT, tests require the use of expensive and limited resources, that possess both monetary and organizational constraints.
Thus, it is necessary when scheduling an IOT campaign to ensure that the generated procedure schedule encompasses both monetary and organizational implications.
The third fitness function, denoted as $\mathit{fitcost}(\mathcal{S})$, provides a means to evaluate such cost associated with a procedure schedule.
When considering the allocation of resources, particularly antennas for an IOT campaign, there exist critical thresholds below which the cost-effectiveness of dedicating the antenna exclusively to the IOT campaign outweighs the benefits of allocating it for other tasks.
Such thresholds can be determined by various factors such as operational efficiency, resource availability, and opportunity costs.
For instance, above a certain number of allotments per day, it may be more cost-efficient to prioritize the IOT campaign, allocating resources to a slot that spans the entire day and postponing other uses of the antenna to a later date, thereby minimizing overall costs.
Similarly, there may exist thresholds or periods during which scheduling a test procedure may not be desirable.
For instance, scheduling a test procedure outside of working hours may be inconvenient for practitioners and induce extra costs for them.

Given a candidate procedure schedule $\mathcal{S}$, we define the fitness function $\mathit{fitcost}(\mathcal{S})$, as follows:
\begin{equation*}
    \mathit{fitcost}(\mathcal{S}) = \frac{\mathit{cost}(\mathcal{S}) - \mathit{cost}_{\text{min}}}{\mathit{cost}_{\text{max}} - \mathit{cost}_{\text{min}}}
\end{equation*}
where $\mathit{cost}(\mathcal{S})$ is the cost of the candidate schedule $\mathcal{S}$, and $\mathit{cost}_{\text{min}}$ (resp. $\mathit{cost}_{\text{max}}$) is the minimal (resp. maximum) theoretical cost achievable.
We note that $\mathit{cost}(\mathcal{S})$ is a cost function defined internally by the IOT operators and is dependent on the specific operational context.
Additionally, $\mathit{cost}_{\text{min}}$ and $\mathit{cost}_{\text{max}}$ are inferred by IoT operators based on their domain knowledge of what would constitute the best-case and worst-case schedules in theory, for their specific operational context.
In Section~\ref{sec:evaluation}, we provide the exact cost function used in our experiments.

\subsubsection{Constraints}

Considering constraints during the search process helps reduce the search space, making the search process more efficient.
The complexity of scheduling test procedures can result in the generation of infeasible schedule solutions.
The definition of an infeasible schedule solution aligns with the definition of conflicts described in Section~\ref{sec:conflicts}, meaning that a schedule is considered infeasible if it contains at least one conflict among the test procedures it includes.
By eliminating infeasible solutions during the search process, our approach can focus on viable solutions, reducing computational time and resources.

Various techniques have been proposed in the literature to handle constraints and infeasible solutions, such as death penalties~\cite{Schwefel1981}, static penalties~\cite{HomaifarQL1994, GenC1999}, repair algorithms~\cite{LiepinsGV1990, GenC1999}, or constraints as objectives~\cite{Coello2000}.
In our approach, we apply penalties to infeasible solutions, as the Niched-Penalty approach~\cite{DebA1999} handles infeasible solutions.
First, our approach measures the degree of infeasibility of a solution as follows:
\begin{equation*}
    g(\mathcal{S}, \mathcal{T}) = \sum_{\substack{(\tau_i, \tau_j)  \in \mathcal{S}^2 \\ \tau_i \neq \tau_j}} \xi_{\tau_1, \tau_2}
\end{equation*}
Subsequently, a penalty is applied to the fitness of $\mathcal{S}$ if $g(\mathcal{S}, \mathcal{T}) > 0$, as follows:
\begin{equation*}
    F(\mathcal{S}) = \begin{cases}
        f(\mathcal{S}) & \text{if $g(\mathcal{S}, \mathcal{T}) \le 0$} \\
        f_{\text{max}} + g(\mathcal{S}, \mathcal{T}) & \text{otherwise}
    \end{cases}
\end{equation*}
where, for brevity, $F(\mathcal{S})$ represents the fitness vector of the candidate schedule $\mathcal{S}$, with each dimension corresponding to a fitness function.
The term $f(\mathcal{S})$ denotes the fitness value for $\mathcal{S}$, and $f_{\text{max}}$ is the maximum fitness value among all feasible solutions in the population.

\subsubsection{Computational search}

\begin{algorithm}[tph]
\caption{An algorithm for selecting the near-optimal procedure schedules, based on NSGA-III}
\label{alg:search_algorithm}
\begin{algorithmic}[1]
    \Input
    \Statex $\mathcal{T}$: possible test procedures
    \Statex $n_p$: size of the population and the archive
    \Statex $n_r$: number of reference points
    \Statex $\mu_c$: crossover probability
    \Statex $\mu_m$: mutation probability

    \Output
    \Statex $\mathcal{\kern-1pt N\kern-1pt S}$: near-optimal procedure schedules
    \Statex

    \Comment{generate the initial population}
    \State $\mathbf{P} \gets \emptyset$
    \Repeat
        \State $I \gets \Call{GenerateSchedule}{\mathcal{T}}$
        \State $\mathbf{P} \gets \mathbf{P} \cup I$
    \Until{$\card{\mathbf{P}} = n_p$}
    \Comment{create an empty archive and reference points}
    \State $\mathbf{P}_\alpha \gets \emptyset$

    \State $\mathbf{R} \gets \Call{GenerateReferencePoints}{n_r}$
    
    \Repeat
        \Comment{assess the fitness of each individual}
        \ForEach{$I \in \mathbf{P}$}
            \If{$g(I, \mathcal{T}) = 0$}
                \State $f_1(I) = \mathit{fituse}(I)$
                \State $f_2(I) = \mathit{fitfrag}(I)$
                \State $f_3(I) = \mathit{fitcost}(I)$
\Else
                \State $f_1(I) = f_{1,\text{max}} + g(I, \mathcal{T})$
                \State $f_2(I) = f_{2,\text{max}} + g(I, \mathcal{T})$
                \State $f_3(I) = f_{3,\text{max}} + g(I, \mathcal{T})$
            \EndIf
        \EndFor

        \State $\mathbf{P}' \gets \Call{AssociateWithReferencePoints}{\mathbf{P}, \mathbf{R}}$

        \Comment{update the archive}
        \State $\mathbf{P}_\alpha \gets \mathbf{P}_\alpha \cup \mathbf{P}'$
        
        \State $\Call{ComputeFrontRanks}{\mathbf{P}_\alpha}$
        \State $\Call{ComputeSparsities}{\mathbf{P}_\alpha}$
        \State $\mathbf{P}_\alpha \gets \Call{SelectArchive}{\mathbf{P}_\alpha, n_p}$
        \Comment{update the Pareto front}
        \State $\mathcal{\kern-1pt N\kern-1pt S} \gets \Call{ParetoFront}{\mathbf{P}_\alpha}$
        
        \Comment{create a new population}
        \State $\mathbf{P} \gets \Call{Breed}{\mathbf{P}_\alpha, n_p, \mu_c, \mu_m}$
    \Until{$\mathcal{\kern-1pt N\kern-1pt S}$ is the ideal Pareto front \textbf{or} the algorithm run out of time}
    
    \State \Return $\mathcal{\kern-1pt N\kern-1pt S}$
\end{algorithmic}
\end{algorithm}

We use the Non-dominated Sorting Genetic Algorithm version 3 (NSGA-III)~\cite{DebJ2014:NSGA-III} to find (near-)optimal schedules of IOT test procedures, as shown in Algorithm~\ref{alg:search_algorithm}.
The NSGA-III algorithm has been successfully applied to several software engineering problems~\cite{MkaouerKBSDO2014, ArrietaWMAES2019}.
The algorithm first generates a set of candidate procedure schedules $\mathbf{P}$ (lines~2-6) and then evolves the population iteratively until finding the best non-dominated schedules (Pareto front) or exhausting the time budget (lines~8-33).
In each iteration, the algorithm first assesses the fitness of the individuals $I \in \mathbf{P}$ using the fitness functions (lines~11-22), and applies a penalty if required.
Calculating the fitness of the individual allows the algorithm to find which candidate schedule to keep in the archive and compute the Pareto front (lines~23-30).
Subsequently, based on the archive and reference points, the algorithm breeds a new population $\mathbf{P}$ (line~32) using the following genetic operators:
(1)~\emph{Selection} chooses the candidates to be selected for reproduction by leveraging a \emph{binary tournament} selection technique~\cite{Luke2013:Metaheuristics};
(2)~\emph{Crossover} generates offspring from two candidate schedules, using a one-point crossover technique~\cite{Luke2013:Metaheuristics};
(3)~\emph{Mutation} introduces diversity in the offspring by modifying some of the test procedures of the offspring according to a mutation rate, and a specific strategy.
Below, we further describe our crossover and mutation operators.

\paragraph{Crossover.}
Our crossover method employs a \emph{one-point crossover} operator~\cite{Luke2013:Metaheuristics}.
Specifically, given two parent candidate schedules $\mathcal{S}^l$ and $\mathcal{S}^r$, each containing IOT test procedures $\{\tau_1^l, \tau_2^l \ldots, \tau_n^l \}$ and $\{\tau_1^r, \tau_2^r \ldots, \tau_n^r\}$ respectively, the crossover operator randomly selects a crossover point $i$.
It then generates two offspring by swapping some test procedures between the parents based on $i$, resulting in $\{\tau_1^r, \ldots, \tau_i^r, \tau_{i+1}^l, \ldots$, $\tau_n^l\}$ and $\{\tau_1^l, \ldots, \tau_i^l, \tau_{i+1}^r, \ldots$, $\tau_n^r\}$.

We note that the resulting child schedules might become infeasible after such a crossover operation.
However, these infeasible schedules are managed through the constraint handling technique detailed previously (i.e., such schedules are inflicted with a penalty).

\paragraph{Mutation.}
Our mutation method is applied to the candidate procedure schedules generated by the crossover operation with a probability $p_{mut}$.
The test procedures to be modified in the selected candidates are chosen using a uniform-mutation operator~\cite{Luke2013:Metaheuristics}.
Algorithm~\ref{alg:mutation} precisely describes our mutation method, which takes as input a set $\mathcal{T}_e$ of possible test procedures for the mutation, the candidate schedule $\mathcal{S}$, the maximum probability of selecting a non-conflicting test procedure $p_\textit{nc}^\text{max}$, and the minimum probability of selecting a non-conflicting test procedure $p_\textit{nc}^\text{min}$.

\begin{algorithm}
\caption{Mutation algorithm}
\label{alg:mutation}
\begin{algorithmic}[1]
    \Input
    \Statex $\mathcal{T}_{e}$: eligible test procedures for mutation
    \Statex $\mathcal{S}$: candidate solution
    \Statex $p_\textit{nc}^\text{max}$: maximum probability of selecting a non-conflicting test procedure
    \Statex $p_\textit{nc}^\text{min}$: minimum probability of selecting a non-conflicting test procedure

    \Output
    \Statex $\mathcal{S}_{mut}$: mutated candidate schedule
    \Statex

    \Comment{initialize the probabilities of selecting a conflicting or non-conflicting test procedure}
    \State $r = 1 - \frac{\Xi(\mathcal{S})}{\card{\mathcal{S}}}$
    \State $P_{\text{non-conflicting}} \gets p_\textit{nc}^\text{min} + r \times (p_\textit{nc}^\text{max} - p_\textit{nc}^\text{min})$
    \State $P_{\text{conflicting}} \gets 1 - P_{\text{non-conflicting}}$
    
    \Comment{assign probabilities to each $\tau \in \mathcal{T}_e$}
    \For{each $\tau \in \mathcal{T}_e$}
        \If{$\tau$ conflicts with $\mathcal{S}$}
            \State $P(\tau) \gets P_{\text{conflicting}}$
        \Else
            \State $P(\tau) \gets P_{\text{non-conflicting}}$
        \EndIf
    \EndFor
    
    \Comment{normalize the probabilities}
    \State $P_{\text{total}} \gets \sum_{\tau \in \mathcal{T}_e} P(\tau)$
    \For{each $\tau \in \mathcal{T}_e$}
        \State $P(\tau) \gets \frac{P(\tau)}{P_{\text{total}}}$
    \EndFor
    
    \Comment{select a test procedure based on the probabilities}
    \State $\tau_{\text{selected}} \gets$ select a test procedure from $\mathcal{T}_e$ using the probabilities $P(\tau)$
    
    \Comment{mutate the candidate schedule}
    \State $\mathcal{S}_{mut} \gets$ mutate $\mathcal{S}$ using $\tau_{\text{selected}}$

\State \Return $\mathcal{S}_{mut}$
\end{algorithmic}
\end{algorithm}

Algorithm~\ref{alg:mutation} initializes the probabilities of selecting a conflicting or non-conflicting test procedure by calculating the ratio $r$ (line~2), which represents the proportion of non-conflicting test procedures in the candidate schedule $\mathcal{S}$, as $r = 1 - \frac{\Xi(\mathcal{S})}{|\mathcal{S}|}$, where $\Xi(\mathcal{S})$ is the number of conflicts in $\mathcal{S}$ and $\card{\mathcal{S}}$ is the total number of test procedures in $\mathcal{S}$.
The probability of selecting a non-conflicting test procedure, $P_{\text{non-conflicting}}$, is computed using linear interpolation between $p_\textit{nc}^\text{min}$ and $p_\textit{nc}^\text{max}$ as $P_{\text{non-conflicting}} = p_\textit{nc}^\text{min} + r \times (p_\textit{nc}^\text{max} - p_\textit{nc}^\text{min})$ (line~3).
The probability of selecting a conflicting test procedure, $P_{\text{conflicting}}$, is then calculated as the complement of $P_{\text{non-conflicting}}$, such that $P_{\text{conflicting}} = 1 - P_{\text{non-conflicting}}$ (line~4)

Next, the algorithm assigns probabilities to each test procedure $\tau \in \mathcal{T}_e$ based on whether it conflicts with the candidate schedule $\mathcal{S}$ (lines~6-12).
If $\tau$ conflicts with $\mathcal{S}$, it is assigned $P(\tau) = P_{\text{conflicting}}$; otherwise, it is assigned $P(\tau) = P_{\text{non-conflicting}}$.
The assigned probabilities are then normalized to ensure they sum to 1 (lines 14-17).
The total probability $P_{\text{total}}$ is computed as the sum of all assigned probabilities, and each probability $P(\tau)$ is normalized by dividing it by $P_{\text{total}}$: $P(\tau) = \frac{P(\tau)}{P_{\text{total}}}$.
Algorithm~\ref{alg:mutation} then selects a test procedure $\tau_{\text{selected}}$ from $\mathcal{T}_e$ using the probability $P(\tau)$ (line~19).
Finally, the candidate schedule $\mathcal{S}$ is mutated using the selected test procedure $\tau_{\text{selected}}$ to obtain the mutated candidate schedule $\mathcal{S}_{mut}$ (line~21), which is then returned as the output (line~22).
 \section{Empirical Evaluation}
\label{sec:evaluation}

This section empirically evaluates our approach using real IOT data obtained from SES Techcom.
Our approach implementation and experiment results are available online~\cite{Artifact-IST}. 

\subsection{Research Questions (RQs)}
\label{sec:research-questions}

\textbf{RQ1 (sanity check).}
\emph{How does our search-based IOT scheduling approach perform compared to random search?}
This research question serves as a fundamental evaluation to validate the effectiveness of our search-based approach~\cite{ArcuriB2014, HarmanMZ2012}.
A well-designed search-based approach is expected to outperform a simple random-search significantly.
If it does not, it would imply that the search process is unnecessary.

\textbf{RQ2 (comparison).}
\emph{How does our approach compare to other search-based techniques?}
We compare our approach to an Ant Colony Optimization (ACO) approach, which is a well-established technique for solving complex scheduling problems.
This comparison aims to demonstrate the effectiveness and efficiency of our approach, by contrasting significantly different search techniques to identify a promising direction.
Thus, RQ2 evaluates the quality of the generated schedules generated by our approach against those produced by the ACO approach.

\textbf{RQ3 (usefulness).}
\emph{How do the schedules generated by our approach compare with the ones generated by engineers?}
To validate the usefulness of our approach, it is crucial to demonstrate that the schedules generated by our approach offer a significant improvement over those manually constructed by experienced engineers.
This research question is critical, as it allows us to provide empirical evidence on the advantages of employing an automated approach compared to traditional manual methods, thereby justifying the need for our approach.
Thus, RQ3 evaluates the quality of the generated schedules over the schedules manually constructed by IOT engineers at SES Techcom.

\subsection{Industrial Study Subject}
\label{sec:industry-subject}

We evaluate our approach on a representative case study system from SES Techcom, specifically, on an IOT campaign for the European Global Navigation Satellite System (GNSS), Galileo.
Such an IOT campaign represents, as discussed in Section~\ref{sec:background}, a type of operational acceptance testing.

Our evaluation relies on a realistic configuration employed in an IOT campaign for the Galileo constellation.
The Galileo constellation orbits the Earth across three distinct Medium Earth Orbit (MEO) planes.
In our experiments, we include twenty-four Galileo satellites as our System Under Test (SUT), where each satellite is denoted throughout the evaluation as $s_1, s_2, \dots, s_{24}$.

During the IOT campaign, SES Techcom conducts SQM tests for each active satellite, complemented by RIOT tests on six of the twenty-four satellites.
SQM test procedures are scheduled to coincide with the maximum elevation pass of each satellite. The tests can either start at, end at, or be centered around the maximum elevation point of the satellite, providing three distinct time slots for a single satellite pass.
In contrast, the RIOT test procedures span the entirety of a satellite pass, starting and ending within a five-degree range of start and end elevation.
Additionally, SES Techcom employs a single antenna for the execution of these test procedures throughout an IOT campaign.

In this IOT campaign, the duration is a maximum of two weeks, between, for example, the 1st of October 2024, and the 14th of October 2024 (denoted $t_1$ and $t_2$, respectively).
To initialize our scheduling approach, we first generate a set of satellite passes at the location of the antenna between the 1st of October 2024  and the 14th of October 2024, $\Gamma_r(t_1, t_2) = \bigcup_{i=1}^{24}\Gamma_r^{s_i}(t_1, t_2)$.
A set of test procedures $\mathcal{T}$ is then defined from $\Gamma_r(t_1, t_2)$, so that for each $\alpha \in \Gamma_r(t_1, t_2)$ an SQM test procedure is associated to it.
In addition, for each $\alpha \in \Gamma_r(t_1, t_2)$ if $\alpha$ belongs to $s_1$, $s_2$, ..., $s_6$, and that $\theta_{start}, \theta_{end} \in \alpha, \theta_{start} \leq 5^\circ, \theta_{end} \leq 5^\circ$, an RIOT test procedure is associated to it.
We then generate the conflict graph of the test procedures, as per its definition (see Section~\ref{sec:conflicts}). 

To run our search-based scheduling approach, as mentioned in Section~\ref{sec:schedule optimization}, a slot scheduling algorithm is required.
In our experiments, we employ Algorithm~\ref{alg:cs-slotting} to create a slot schedule for a given procedure schedule.
\begin{algorithm}[t]
\caption{An algorithm for scheduling slots.}
\label{alg:cs-slotting}
\begin{algorithmic}[1]
    \Input
    \Statex $\mathcal{S}$: procedure schedule to generate a slot schedule

    \Output
    \Statex $\mathcal{Q}$: slot schedule
    \Statex

    \Comment{generate an initial set of slots}
    \State $\mathcal{Q} \gets \emptyset$
    \ForEach {$s \in \mathcal{S}$}
        \State $\mathcal{Q} \gets \mathcal{Q} \cup \{\Call{GenerateSlot}{s}\}$
    \EndFor
    \Comment{sanitize the slot schedule}
    \State $\Call{CombineOverlappingSlots}{\mathcal{Q}}$
    \State $\Call{ConsolidateSlots}{\mathcal{Q}}$

    \State \Return $\mathcal{Q}$
\end{algorithmic}
\end{algorithm}
From lines 1 to 5, an initial set of slots is generated by the algorithm for each test procedure present in the procedure schedule.
Specifically, the time slots in our experiments start at the beginning of an hour, a quarter past the hour, half-past the hour, or three-quarters past the hour.
Additionally, the duration of the time slot must be an integer multiple of one hour, ensuring the completion of the corresponding test procedure.
From lines 6 to 7, several sanitization steps are conducted.
First, slots that overlap with one another are merged so they form a single slot.
Then, slots are consolidated according to the following internal policy at SES Techcom: if more than six hours in twenty-four hours are dedicated to performing IOT procedures, this entire period should dedicated solely to performing IOT procedures, i.e., all those slots are replaced by a single slot that spans a length of twenty-four hours.
Finally, the resulting slot schedule is returned.

We define the $\mathit{cost}$ of a slot schedule $\mathcal{Q}$ as follows:
\begin{equation*}
    \mathit{cost}(\mathcal{Q}) = \sum_{q \in \mathcal{Q}}\mathit{cost}(q)
\end{equation*}
where $q$ is a slot in the slot schedule $\mathcal{Q}$ and $\mathit{cost}(q)$ is defined as
\begin{equation*}
    \mathit{cost}(q) =
    \begin{cases}
        \frac{\Call{span}{q}}{60} \times 456 & \text{if $\Call{span}{q} < 24h$} \\
        3561 & \text{otherwise}
    \end{cases}
\end{equation*}
The cost function $\mathit{cost}(q)$ is determined by the span of the slot $q$.
If the span is less than 24 hours, the cost is calculated as $\frac{\Call{span}{q}}{60} \times 456$, where the span of $q$ (in minutes) is converted to hours ($\frac{\Call{span}{q}}{60}$) and then multiplied by 456.
If the span is 24 hours or more, the cost is fixed at 3561.

\subsection{Experimental Setup}
\label{sec:exprimental setup}

For the evaluation of the research questions, we implemented the following three approaches:
(1)~GSC: Our approach using NSGA-III, described in Section~\ref{sec:approach},
(2)~ACO: An approach using Ant Colony Optimization~\cite{DorigoMC1996},
and (3)~RS: A Random Search approach.

\noindent\textbf{EXP1.}
To answer RQ1, we conduct a comparative analysis of GSC against RS.
We implemented GSC, an IOT scheduling tool based on our approach, described in Section~\ref{sec:approach}.
Additionally, we implemented RS, a random search approach that creates IOT schedules for a given set of satellite passes.
RS creates procedure schedules by randomly selecting $n$ test procedures $\tau \in \mathcal{T}$ (see the definition of $\mathcal{T}$ in Section~\ref{sec:approach}).
To perform our comparison, we evaluate the results of GSC and RS by comparing the resulting fitness values of the solutions.

To further measure the effectiveness of GSC, we use the three quality indicators as described below, following established guidelines found in existing literature~\cite{WangAYLL2016}.
As the optimal solution for our search problem is unknown apriori, we construct a reference Pareto front by combining all non-dominated solutions obtained from each execution of the compared approaches.
We then assess the \textit{Generational Distance} (GD), a metric which measures the Euclidean distance between a specific solution and the nearest solution on a reference Pareto front~\cite{VanVeldhuizenL1998}.
The lower the GD metric, the closer a solution is to the optimal output.
We then assess the \textit{Spread} (SP), a measure of the distance between each point of the Pareto front~\cite{DebPAM2002:NSGA-II}.
The lower the SP metric, the more the non-dominated solutions are spread across the Pareto front, showing higher diversity.
Finally, we assess the \textit{Hypervolume} (HV), a quality indicator which represents the size of the space covered by a search algorithm~\cite{ZitzlerT1998}.
The higher the HV metric, the more space the Pareto front covers, showing better coverage performance.

\noindent\textbf{EXP2.}
To answer RQ2, in this experiment, we compare our approach GSC against an alternative approach ACO relying on Ant Colony Optimization.
The ant colony algorithm is an optimization algorithm which takes inspiration from the foraging behavior of ant colonies.
It has been used in several combinatorial optimization problems, such as knapsack problems~\cite{KongTK2008} and routing problems~\cite{MontemanniGRD2005, BellM2004}.
Notably, the ant colony algorithm has been extensively used in the literature to solve the multi-satellite control resource scheduling problem (MSCRSP)~\cite{ZhangZK2011, GaoWZ2013, WuLMQ2013, ZhangZF2014, ZhangHZ2018}, a combinatorial problem in the satellite domain similar to our IOT scheduling problem.
Generally, these methods are based on the Max–min Ant System (MMAS) proposed by~\citet{StutzleH2000}.
Hence, we select it to create an alternative approach for comparison with GSC.
We describe below our ACO approach, which relies on MMAS.

\textit{Construction of the optimization problem:}
In MMAS~\cite{StutzleH2000}, our IOT scheduling problem is represented as a graph $G = (A, E, T, H)$, where $A$ is the set of test procedures, $E$ is the set of edges that represent execution orders between test procedures, and $T$ and $H$ are the vectors that represent the trail of pheromone and heuristic information, respectively, both associated with the edges in $E$.
The heuristic information represents the problem-specific knowledge that guides ants toward more promising solutions by influencing their path choices based on the desirability of moving to a particular node.
In our problem, the heuristic information is represented by a matrix $H$ of dimension $\card{A} \times \card{A}$, where each element indicates the desirability of transitioning from one test procedure to another.
This desirability is defined by the following function:
\begin{equation*}
\eta(\tau_i, \tau_j) = 
\begin{cases} 
    0 & \text{if } \Xi(\mathcal{S}\mid\tau_j) > 1 \\
    \frac{\Delta(F(\mathcal{S}\mid\tau_j))}{\max\Delta(F(\mathcal{S}))} & \text{otherwise}
\end{cases}
\end{equation*}
where $\tau_i$ and $\tau_j$ represent the current test procedure and the candidate test procedure, respectively.
The term $\Delta(F(\mathcal{S}\mid\tau_j))$ denotes the improvement in the fitness of the schedule $\mathcal{S}$ if $\tau_j$ is included, and $\Xi(\mathcal{S}\mid\tau_j)$ indicates the conflicts in the schedule $\mathcal{S}$ if $\tau_j$ is included.

Solutions to our IOT scheduling problem are represented as paths on the graph $G$.
Ants create candidate solutions by taking randomized walks on the fully constructed graph $G$, guided by the pheromone trail intensity and current heuristic information on the edges.
The conflicts definition $\Xi$ is applied to ensure the ants do not form infeasible solutions while moving between vertices.
After the ants complete their walk, the pheromone trails are updated.

\textit{Initializing the pheromone values:}
Firstly, the approach initializes all the pheromone values to the maximum value.

\textit{Constructing a solution:}
$m$ ants are placed on randomly selected test procedures.
Each ant uses a probabilistic decision-making rule, known as the random proportional rule, to determine which test procedure to select next at each step of the process.
If ant $k$ is at node $i$ during iteration $t$, it will select the next node $j$ based on a certain probability.
\begin{equation*}
    p_{ij}^k(t) = \begin{cases}
        \frac{[\tau_{ij}(t)]^\alpha \cdot [\eta_{ij}]^\beta}{\sum_{l\in N_i^k}[\tau_{il}]^\alpha \cdot [\eta_{il}]^\beta}, & \text{if $j \in N_i^k$} \\
        0,              & \text{otherwise}
    \end{cases}
\end{equation*}
where $\eta_{ij}$ is a heuristic value, $\tau_{ij}$ is the pheromone trail value, $\alpha$ and $\beta$ are two parameters which determine the relative influence of the pheromone trail and the heuristic information, and $N_i^k$ is the feasible procedures of the $k$-th ant when beginning at node $i$.

\textit{Updating the pheromone:}
After all the ants have completed an iteration, the pheromone trails are updated as follows:
\begin{equation*}
    \tau_{ij}(t+1) = \left[(1-\rho)\tau_{ij}(t) + \Delta\tau_{ij}^{\text{best}}(t)\right]_{\tau_{\text{min}}}^{\tau_{\text{max}}}
\end{equation*}
where $\rho$ is the evaporation rate, comprised between 0 and 1, $\Delta\tau_{ij}^{\text{best}}$ is the quantity of pheromone laid by the $k$-th ant on the path visited.
$\Delta\tau_{ij}^{\text{best}}$ is defined as follow:
\begin{equation*}
    \Delta\tau_{ij}^{\text{best}}(t) = \begin{cases}
        F(s^{\text{best}}(t)), & \text{if $(i, j) \in s^{\text{best}}(t)$} \\
        0,                     & otherwise
    \end{cases}
\end{equation*}
where $s^{\text{best}}$ is the best-so-far solution, which is the best solution found during the current iteration or the global-best solution found by the $k$-th ant.
$F$ represents the fitness function that is used to assess the quality of the solution found by the ant.
In ACO, we define it as
\begin{equation*}
    F(\mathcal{S}) = \frac{\textit{fituse}(\mathcal{S}) + \textit{fitfrag}(\mathcal{S}) + \textit{fitcost}(\mathcal{S})}{3}
\end{equation*}

\noindent\textbf{EXP3.}
To answer RQ3, we evaluate the usefulness of our approach, GSC, over IOT schedules generated by SES Techcom's IOT engineers.
We obtained the test configurations and the IOT schedules created by the engineers for the previous IOT campaign conducted by SES Techcom on November 7, 2023.
Additionally, we acquired the original satellite passes that engineers used to develop their schedules.

For the comparison, we executed GSC 50 times using the same configurations and satellite passes as those used by the engineers.
In EXP3, we compared the cost, span, and number of slots required to execute the schedules generated by GSC and those created manually by the engineers.
Furthermore, we assessed the average execution time for GSC to generate a set of schedules and compared it to the time reported by engineers to create their schedules.

\subsection{Parameters Setting}
\label{sec:parameters-tuning}

To perform EXP1 and EXP3, we configured the hyperparameters of GSC as follows: the population size is set to 200, the mutation rate to 0.2, and the crossover rate to 0.8.
Additionally, reference directions for half the population size are generated according to the \emph{Riesz s-energy} principle~\cite{BlankDDBS2021}.
These parameter values follow the guidelines presented in the literature~\cite{ArcuriF2011}.
The termination condition of the approach is set according to the number of fitness function evaluations performed.
To identify the appropriate number of fitness evaluations, we conducted 25 initial experiments, each terminating at 150,000 fitness functions' evaluations, and monitored the evolution of the SP and HV metrics.
We observed that, on average, after 50,000 fitness evaluations, the metrics did not show notable improvement.
Therefore, we set the stopping conditions of our approach to 50,000 fitness function evaluations.
Additionally, we constrain the approaches to a maximum time limit of one hour to ensure that no single run exceeds a reasonable duration.

Regarding ACO used in EXP2, we selected the hyperparmeters based on guidelines in the literature~\cite{DorigoMC1996}.
Specifically, we set the number of ants to 50, the pheromone importance factor $\alpha$ to 1, the heuristic importance factor $\beta$ to 1, the evaporation rate $\rho$ to 0.5, and the pheromone deposit amount to 100.

Nonetheless, we note that those hyperparameters could be further tuned to increase the performance of our approach, however, the results obtained with the described values are sufficient and convincingly support our analysis.
Therefore we do not report further optimizations of those parameters.

\subsection{Experiment Results}
\label{sec:results}

\noindent\textbf{RQ1.}
\begin{table}[t]
    \caption{Comparing GSC and RS Pareto front using the Hypervolume (HV), Spread (SP), and Generational Distance (GD) quality indicators.} 
    \label{tbl:rq1:quality}
    \small
    \centering
    \begin{tabularx}{\linewidth}{lYYYY}
        \toprule
         Metric & p-value & $\hat{A}_{12}$ & Mean GSC & Mean RS\\
         \midrule
SP & 5.65e-3  & 1.0 & \textbf{1.50}  & 0.0  \\
            HV & 5.65e-3  & 1.0 & \textbf{0.131} & 0.0 \\
            GD & 1.73e-17 & 1.0 & \textbf{8.49e-3} & 3.42e-1 \\
\bottomrule
    \end{tabularx}
\end{table}
Table~\ref{tbl:rq1:quality} compares the sets of schedules obtained by GSC and RS after 50 runs of each approach, in terms of their quality indicators.
For statistical comparison, we use Mann-Whitney U-test~\cite{Mann47} and Vargha and Delaney’s $\hat{A}_{12}$ effect size~\cite{Vargha00}.
The level of significance ($\alpha$) is set to 0.05.
Two distributions are considered not superior to each other when the value of $\hat{A}_{12}$ is 0.5.
The table indicates that the solutions found by GSC (i.e., its Pareto front) are of significantly better quality than the solutions found from RS.
Indeed, for each comparison, the p-values are lower than 0.05, and the $\hat{A}_{12}$ values indicate large effect size ($\hat{A}_{12} > 0.5$), which supports the hypothesis that the first distribution is significantly greater than the second distribution.

\begin{figure*}
    \centering
    \includegraphics[width=\textwidth]{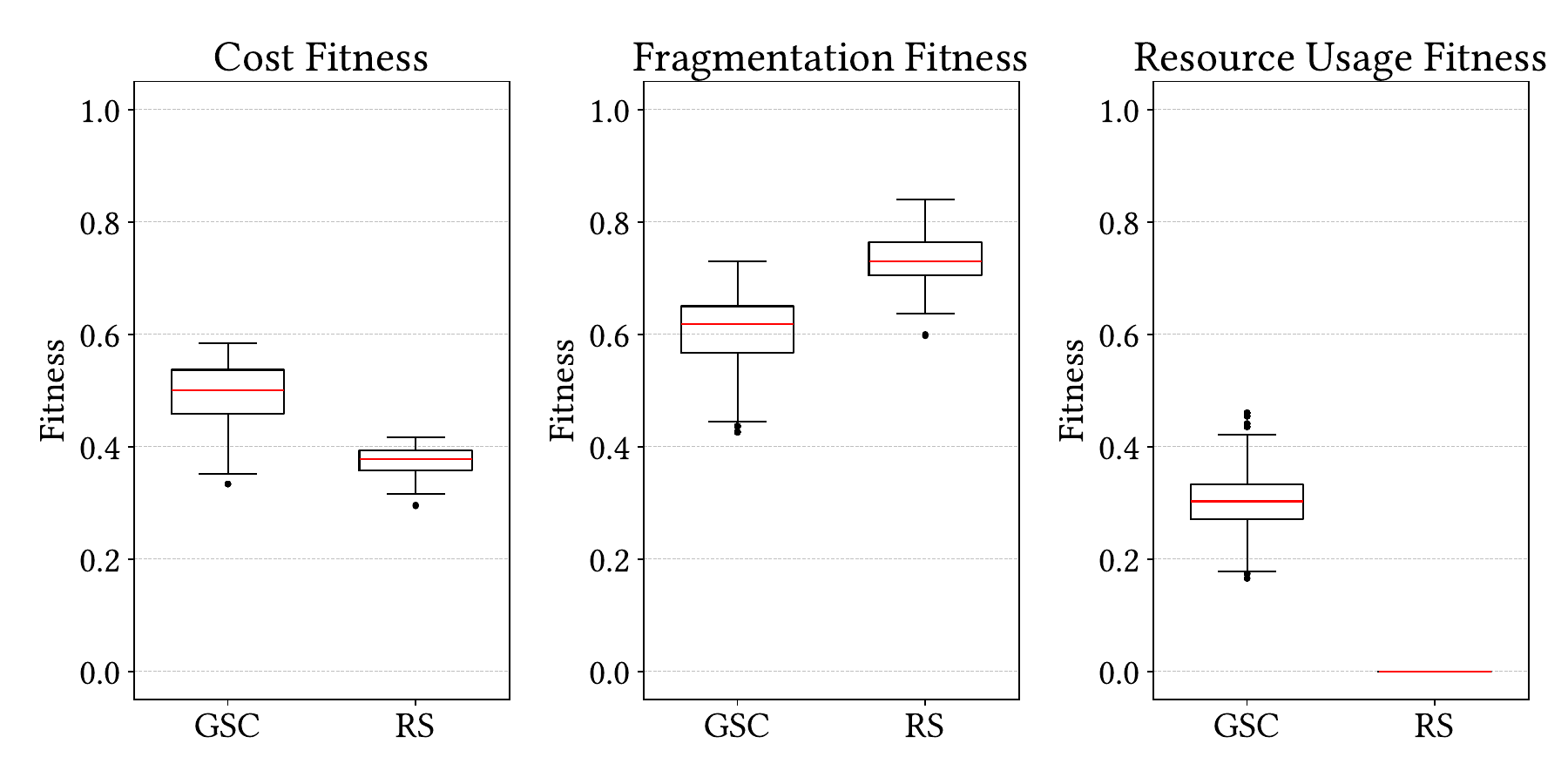}
    \caption{Comparing GSC and RS in terms of \textit{fitcost}, \textit{fitfrag}, and \textit{fituse}. For brevity, we present the fitness values: the higher the fitness value, the better.}
    \label{fig:rq1:fitness}
\end{figure*}
Figure~\ref{fig:rq1:fitness} compares the distributions of the three fitness functions' values (see Section~\ref{sec:schedule optimization}), for each set of solutions obtained after 50 runs of each approach.
The results show that GSC (resp. RS) reaches a fitness of 49.5\% (resp. 37.1\%) for \emph{fitcost}, 60.4\% (resp. 73.1\%) for \emph{fitfrag}, and 30\% (resp. 0\%) for \emph{fituse}.
Those results indicate that in terms of fitness, GSC finds solutions that are more cost-efficient (better \emph{fitcost}) compared to RS, and finds schedules that improve on the test resource usage (better \emph{fituse}).
In terms of \emph{fituse}, we note that the antenna resource usage cannot be calculated when schedules are not feasible which explains why the average fitness value \emph{fituse} is 0 for RS.
This also explains why the second fitness function (\emph{fitfrag}) is higher for RS than for GSC, as when schedules are infeasible, there exists some overlap between the different test procedures, therefore this may tend to a better fragmentation as infeasible test procedure may be very close to each other or overlapping, thus presenting less gap between them.

\begin{figure*}
    \centering
    \includegraphics[width=\textwidth]{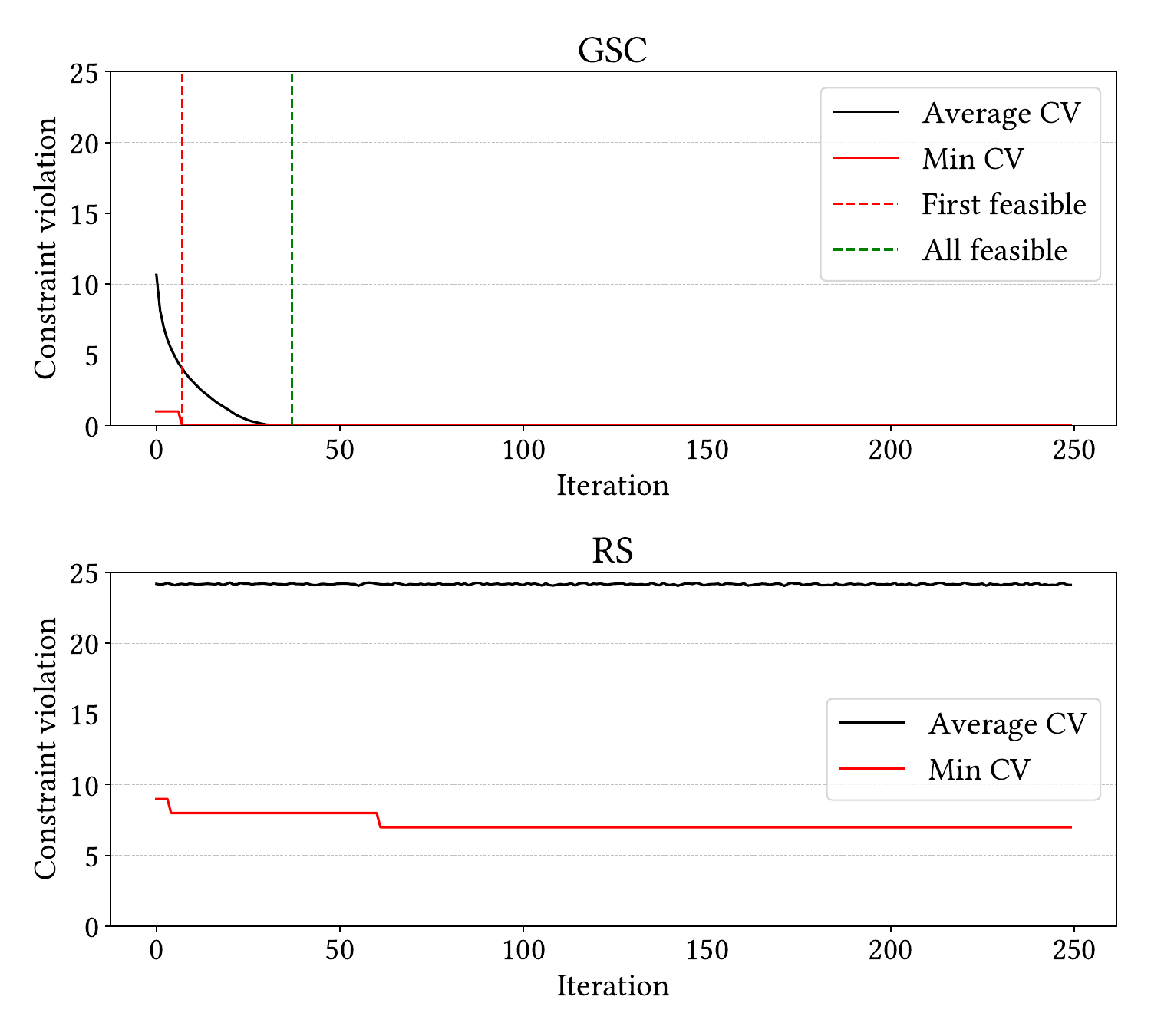}
    \caption{Comparing the progression of the constraint violations between GSC and RS.}
    \label{fig:rq1:cv}
\end{figure*}
To further analyze how GSC and RS handle infeasible schedules, Figure~\ref{fig:rq1:cv} compares the progression of average and minimum constraint violations over 50 runs of GSC and RS.
The results show that on average, GSC can find the first feasible schedule after seven iterations and, on average, all schedules become feasible after 37 iterations.
On the contrary, RS produces schedules with, at best, seven conflicts and the schedules have, on average, 24 conflicts per iteration.

\begin{tcolorbox}[enhanced jigsaw,left=2pt,right=2pt,top=0pt,bottom=0pt]
\emph{The answer to} \textbf{RQ1} \emph{is that} our approach, GSC, significantly outperforms RS in generating schedules for IOT campaigns. 
In particular, RS cannot generate schedules that do not violate the constraints imposed by the IOT scheduling problem.
\end{tcolorbox}

\begin{figure*}
    \centering
    \includegraphics[width=\linewidth]{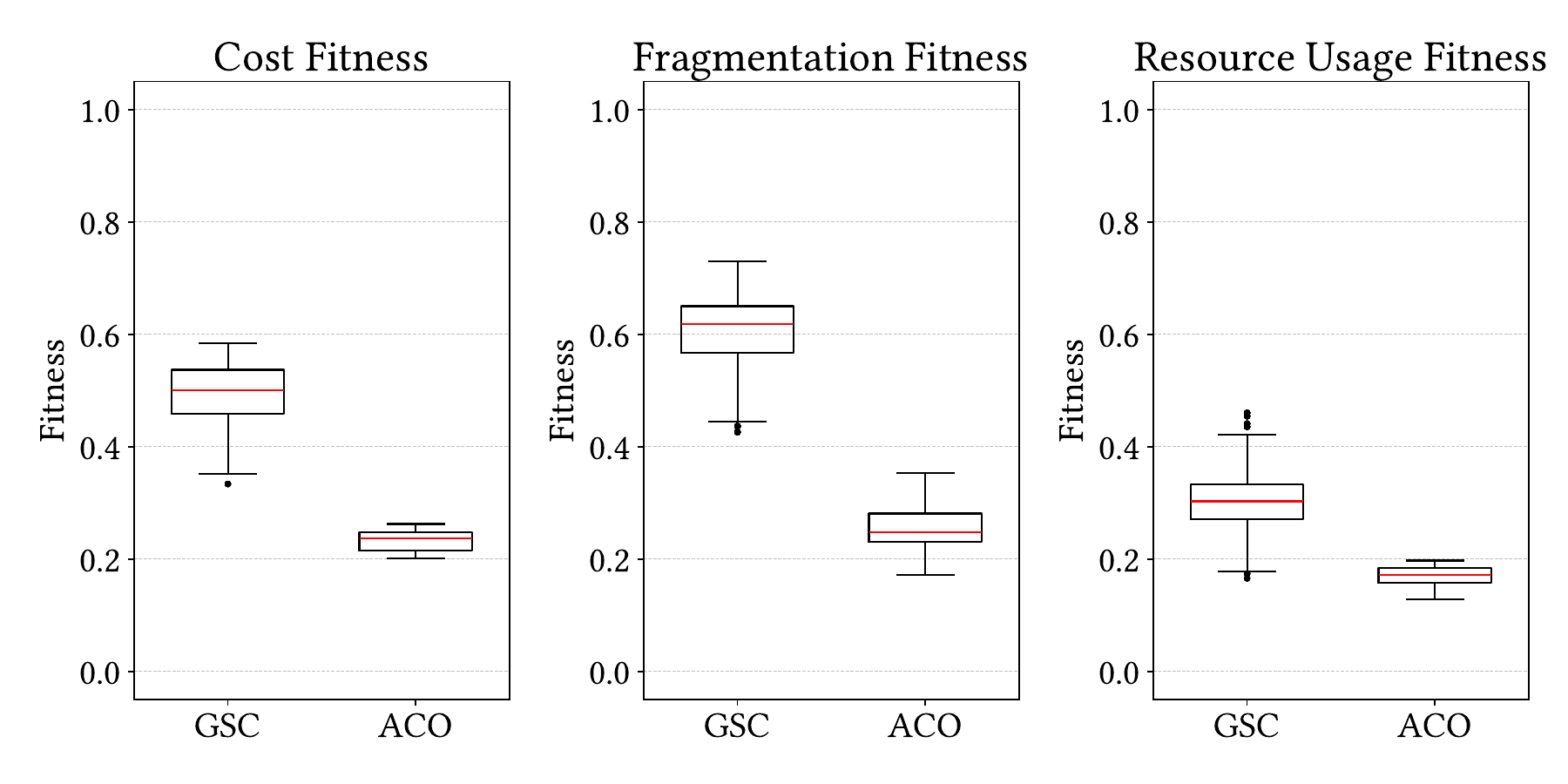}
    \caption{Comparing GSC and ACO in terms of \textit{fitcost}, \textit{fitfrag}, and \textit{fituse}. For brevity, we present the fitness values: the higher the fitness value, the better.}
    \label{fig:rq2:fitness}
\end{figure*}
\noindent\textbf{RQ2.}
Figure~\ref{fig:rq2:fitness} compares the non-dominated solutions obtained after 50 runs of ACO described in EXP2 and our approach GSC.
The figure shows that for the fitness functions \textit{fitcost}, \textit{fitfrag}, and \textit{fituse}, GSC obtains, on average, a score of 0.49, 0.60, and 0.30, respectively.
In contrast, for the same fitness functions, ACO obtains, on average, a score of 0.23, 0.25, and 0.17, respectively.

\begin{table}[ht]
    \centering
    \caption{Comparison of the average number of iterations performed and feasible schedules obtained after 50 runs of GSC and ACO.}
    \label{tab:rq2:exec_stats}
    \begin{tabularx}{\linewidth}{Xcc}
        \toprule
        \textbf{Method} & \textbf{Iterations (Average)} & \textbf{\# Feasible Schedules (Average)} \\
        \hline
        GSC & 250 & 38 \\
        ACO & 13 & 1 \\
        \bottomrule
    \end{tabularx}

\end{table}
Additionally, Table~\ref{tab:rq2:exec_stats} compares the number of iteration performed, and number of solutions (i.e., feasible schedules) obtained on average for 50 runs of GSC and ACO.
Each approach run was constrained to one hour to ensure a fair comparison.
The table indicates that GSC, with a population of 200 candidate schedules, performs an average of 250 iterations, equating to 50,000 fitness evaluations.
In contrast, ACO, with a population of 20 ants, performs an average of 13 iterations, resulting in 260 fitness evaluations.
Furthermore, the table shows that GSC yields an average of 38 equally viable schedules, whereas ACO produces a single feasible schedule.

Those results indicate that GSC's ability to perform more fitness evaluations allows it to explore the search space more thoroughly within a limited time frame.
Additionally, by producing a greater number of equally viable solutions, GSC enables practitioners to conduct trade-off analyses.

\begin{tcolorbox}[enhanced jigsaw,left=2pt,right=2pt,top=0pt,bottom=0pt]
\emph{The answer to} \textbf{RQ2} \emph{is that} our approach, being more time-efficient, allows for a more thorough exploration of the search space.
This results in our approach outperforming ACO in terms of the fitness of the solutions found.
Additionally, unlike ACO, our approach produces several equally viable schedules, enabling practitioners to perform trade-off analyses.
\end{tcolorbox}

\begin{figure*}
    \centering
    \includegraphics[width=\textwidth]{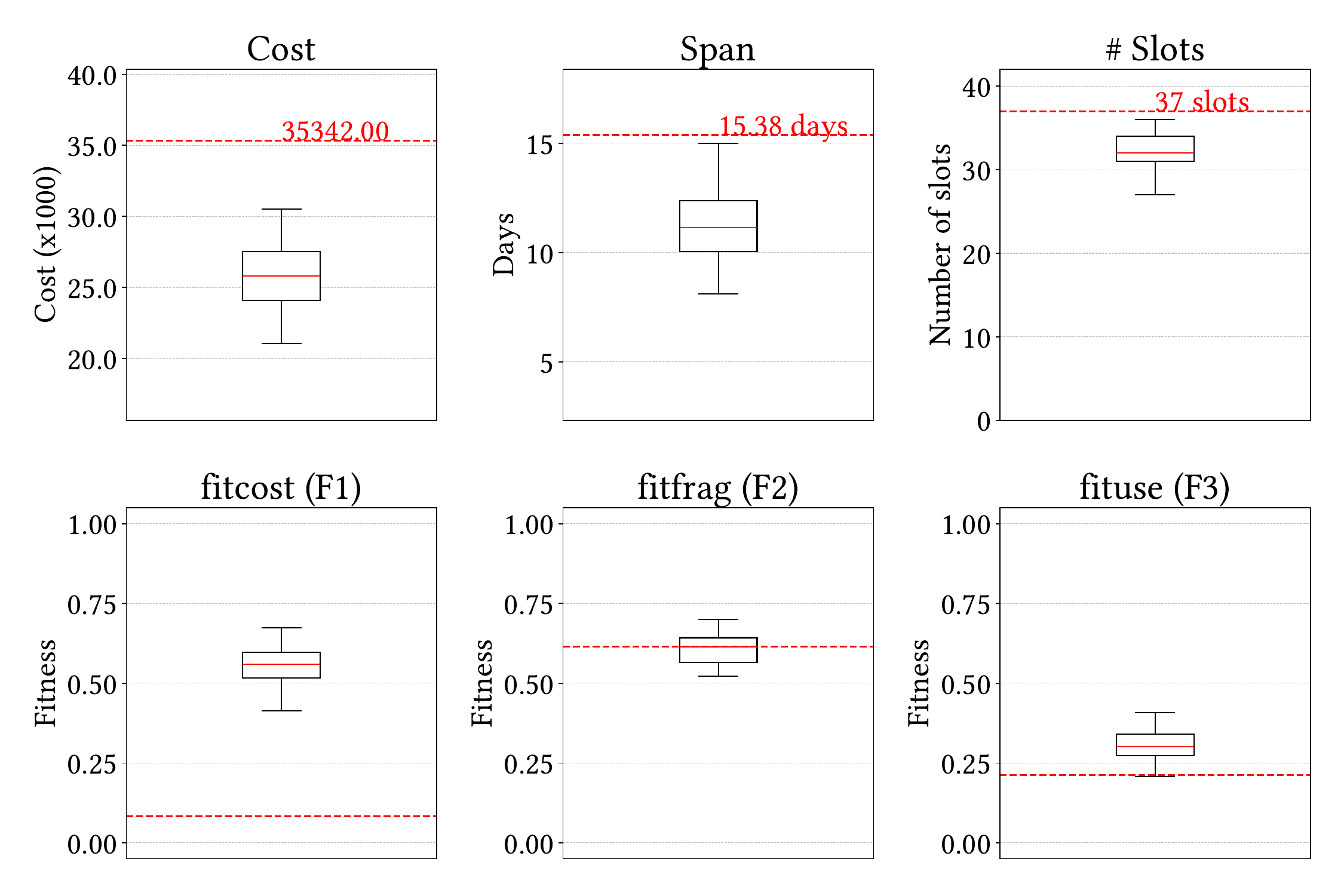}
    \caption{Comparing the cost, span, number of slots, \textit{fitcost}, \textit{fitfrag} and \textit{fituse} of the our GSC against manually crafted schedules. For brevity, we present the fitness values: the higher the fitness value, the better.}
    \label{fig:rq3:fitness}
\end{figure*}
\noindent\textbf{RQ3.}
We compare the schedules generated by our approach GSC with the reference schedule provided by SES Techom (see Section~\ref{sec:exprimental setup}).
Figure~\ref{fig:rq3:fitness} compares the distributions of schedule costs, spans, and the number of slots between the generated schedules and the reference schedule.
Furthermore, Figure~\ref{fig:rq3:fitness} presents the distributions of fitness values obtained from 10 runs of GSC, compared to the calculated fitness values of the reference schedule.
The figure indicates that the average cost for the generated schedules is 25,657, with an average span of 11 days and 32 slots.
In comparison, the reference schedule has an average cost of 35,342, a span of 15.38 days, and 37 slots.
This represents an average reduction of 27\% in cost, 28\% in span, and 13\% in the number of slots compared to the manually crafted schedule.

Furthermore, the figure shows that, for \emph{fitcost}, generated schedules have an average fitness of 53\%, compared to an average of 8.3\% for the reference schedule. 
This represents an increase of approximately 538\% in the average fitness, indicating that the generated schedules are significantly more cost-efficient.
Regarding \emph{fitfrag}, the average fitness is approximately 60\%, compared to 60.9\% for the reference schedule.
This indicates that, on average, our approach is equally effective in minimizing the amount of context switching required from engineers.
Finally, for \emph{fituse}, the average fitness is approximately 29\%, compared to 20.8\% for the reference schedule.
This represents a 39.42\% increase in the efficiency of antenna resource usage for the generated schedules compared to the reference schedule.

\begin{tcolorbox}[enhanced jigsaw,left=2pt,right=2pt,top=0pt,bottom=0pt]
\emph{The answer to} \textbf{RQ3} \emph{is that} the schedules generated by our approach are significantly more cost-efficient, with shorter schedule spans and fewer slots.
They also maintain comparable or better performance in terms of fragmentation and resource usage efficiency compared to those generated by expert engineers.
\end{tcolorbox}

\subsection{Lessons Learned}
\label{sec:lessons-learned}

To further assess the practical usefulness of our approach, we reached out to three IOT engineers at SES Techcom to obtain feedback on our work and discuss possible improvements.
The three engineers are currently working on IOT systems.
One is a senior manager with several years of experience, while the other two are junior engineers currently in charge of IOT projects.
All three engineers have previously handcrafted test schedules for IOT campaigns.

Following the evaluation of GSC, our approach, conducted for RQ1 and RQ3, we provided it to the engineers.
We began by presenting a detailed demonstration, explaining GSC’s usage and the impact of its various parameters.
Subsequently, the engineers were given a period of four hours to familiarize themselves with GSC and utilize it to generate IOT schedules they had previously crafted manually.
After this familiarization period, we gathered their feedback, focusing on GSC’s usability, the quality of the generated schedules, and their perspectives on integrating GSC into their workflow.
Overall, the experts' feedback highlighted three key takeaways from our work.

\noindent\textbf{Efficiency of schedule generation.}
All three engineers acknowledged that the generated schedules are more conveniently arranged than those they thoroughly handcrafted.
This observation aligns with our findings in RQ1 and RQ3, where we concluded that the schedules generated by our approach outperform their handcrafted counterparts across all metrics, including fitness.
Additionally, the engineers highlighted that the ability to generate schedules in an automated manner, is a significant advantage, as it requires only a fraction of the time required for creating such schedules manually, and the automation of the scheduling process reduces the likelihood of human errors.

\noindent\textbf{Multiple schedules enable tradeoff analysis.}
Engineers highlighted that receiving several schedules in a short time frame could significantly impact their operations.
Specifically, they stated that this capability allows for quick adaptation to changing conditions (e.g., bad weather forecast, unavailability of equipment or personnel) during testing and better alignment with the needs of specific test campaigns.
For example, some IOT campaigns require minimum fragmentation, even if it comes at the expense of cost. 
Having multiple equally viable schedules enables trade-off analysis for these varying needs and thus is beneficial.

\noindent\textbf{Room for improvement.}
The senior engineer and one junior engineer noted that while the generated schedules are highly efficient, there is potential for further improvement.
They mentioned that the reconfiguration time overhead between two test procedures is currently fixed but is largely dependent on the time required to reposition the antenna.
This overhead can be reduced by selecting test procedures where the satellite positions at the end of one procedure and the start of the next are relatively close to each other.
Furthermore, they suggested that the schedules could exhibit greater \textit{``diversity''}, by increasing the spread of the final set of generated schedules.
This could enhance the variety of schedules available to the engineers to choose from.
Finally, they proposed adding additional constraints and requirements, such as formalizing the notion of risk associated with a schedule and incorporating it as one of the objective functions.
According to their explanation, the risk of a schedule is mainly determined by the consecutive duration of hardware usage.
This means that executing many tests consecutively over a short period poses less risk than executing them separately over a longer time period.
This notion is already partially covered by our second fitness function, $\textit{fitfrag}$, but formally introducing it as an objective for our approach could be beneficial for practitioners.
They also suggested introducing the concept of priority, suggesting that performing certain test procedures earlier than others may be beneficial.
Therefore, future research could focus on improving our approach to account for those requirements.

\subsection{Threats to Validity}
\label{sec:threats}

\noindent\textbf{Internal validity.}
The primary internal threat to validity is the potential presence of hidden variables that may weaken the relation between the results obtained for the different approaches.
To minimize this impact, we evaluated each approach using the same parameter settings.
Additionally, we disclose all the configurations and share our experimental data to ensure reproducibility.

\noindent\textbf{External validity.}
The primary threat to external validity is the potential lack of generalizability of our results to other contexts.
This threat can be further divided into two aspects: (1)~the extent to which our approach can be applied to systems different from our case study, and (2)~whether similar benefits observed in our case study can be replicated in different contexts.
For the first aspect, we thoroughly described the IOT requirements in Section~\ref{sec:requirements and constraints}, which are necessary for our approach to be applicable. 
As long as any IOT context meets these requirements, our approach remains applicable.
These requirements are based on generic IOT campaigns, which are relevant to many satellite systems.
Regarding the second aspect, although our case study was conducted in a representative realistic setting, additional case studies are required to validate our approach.
However, we note that our approach is currently being used by practitioners, providing further confirmation that our approach may be applicable in various IOT contexts.
 \section{Related Work}
\label{sec:related-work}

\noindent\textbf{Operational acceptance testing.}
Most research on acceptance testing focuses on its application in agile software development methodologies (e.g., SCRUM) and User Acceptance Testing (UAT)~\cite{LeungW1997, Finsterwalder2001, DavisV2004, LofflerGG2010, LiskinHKKRS2012, LiebelAF2013}.
However, few works in the literature focus on test case scheduling of Operational Acceptance Testing (OAT)~\cite{AmmannO2016,ShinNSBZ2018}.
A notable work in this field was proposed by~\citet{ShinNSBZ2018}, where they developed a method to automate test case prioritization for Cyber-Physical Systems (CPS) acceptance testing.
Their method accounts for time budget constraints, uncertainty, and hardware damage risks posed by the sequential execution of test cases.
Unlike their method, our work studies OAT in the context of mission-critical satellite systems, which accounts for different factors, such as antenna utilization, operational cost and context switching.

\noindent\textbf{Test case prioritization.}
A research strand closely related to our work is test case prioritization, which has been widely studied in the literature, especially in the topic of software regression testing~\cite{LiHH2007, BusjaegerX2016, CatalM2013, KhatibsyarbiniIJT2018}.
Among these, the research strand that most closely aligned with our work concerns the prioritization of test cases for CPSs.
Many of those techniques consider test execution time as the primary resource to be used for the prioritization process, with other objectives being based on the source code of the software under test.
For example, \citet{ArrietaWSE2016} proposed a weight-based multi-objective search algorithm that prioritizes test cases for configurable CPSs, within each system configuration to optimize the testing process, considering the test execution time and the success rate of the tests at hand.
\citet{WangAYOL2016} proposed a multi-objective search-based approach that prioritizes test cases in CPSs considering the execution time of the test cases as well as the hardware resource requirements.
\citet{ArrietaWSE2019} proposed a search-based approach that prioritizes test cases for CPS product lines, aiming to optimize the testing process cost-efficiently.
This approach focuses on reducing fault detection time, simulation time, and the time required to cover both functional and non-functional requirements.
However, these works prioritize test cases without addressing scheduling challenges, as the tests are executed sequentially without considering resource availability.
Additionally, they do not account for conflicts between candidate test cases, nor the availability of operators, making these approaches not directly applicable to address our IOT scheduling problem.

\noindent\textbf{Resource scheduling for satellite systems.}
In the satellite domain, the research strand related to our work is the satellite control resource scheduling problem (SCRSP), specifically, the sub-problem of Ground Measurement and Control Resource Allocation (GMCRA).
This research strand addresses the allocation of ground resources for satellite control and measurement activities.
For instance, \citet{MarinelliNRS2011} introduced a Lagrangian heuristic algorithm for satellite range scheduling with resource constraints.
By framing the problem as a sequence of maximum weighted independent set problems on interval graphs, \citet{MarinelliNRS2011} applied a Lagrangian relaxation to schedule satellite communication requests from a ground station to the Galileo GNSS constellation.
\citet{ZhangZK2011, GaoWZ2013, WuLMQ2013}, and~\citet{ZhangZK2011, ZhangZF2014, ZhangHZ2018} proposed ant colony optimization-based algorithms which take into account the task interval constraints, resource availability constraints, and satellite constraints to provide efficient scheduling solutions.
However, none of these approaches are directly applicable to our problem because they do not fully address the specific constraints and requirements of scheduling IOT test procedures.

 \section{Conclusions}
\label{sec:conclusions}

In this article, we presented an approach for generating IOT schedules for operational acceptance testing of mission-critical satellite systems.
Our approach, based on a multi-objective search algorithm, allows practitioners to efficiently generate schedules while satisfying objectives related to test resource usage, operational costs, and context switching.
We evaluated our approach using industry IOT data from SES Techcom, specifically for the IOT campaigns of the Galileo GNSS constellation.
Our results indicate that our approach effectively addresses the IOT scheduling problem and outperforms alternative approaches based on random search and ant colony optimization.
The generated schedules show an average cost reduction of 538\% compared to manually created schedules, while maintaining low context switching and improving test resource efficiency by 39.42\%.

Feedback from engineers highlighted additional benefits of our approach, such as reducing human error in test schedule generation through automation, simplifying schedule creation (with implementation taking less than one hour compared to a full workday for manual schedules), and enabling trade-off analysis by providing multiple viable schedules.

Engineers also suggested enhancing our approach by increasing the diversity of generated schedules and incorporating additional constraints, such as risk and priority.
Future work will focus on integrating these aspects and addressing the need for more diversity.
Additionally, we plan to apply our approach to a broader range of study subjects, including multiple antennas and different constellations. 
\section*{Data Availability Statement}
The implementation of our approach and the experiment results are made available online~\cite{Artifact-IST} to facilitate reproducibility and adoption by researchers and practitioners.

\begin{acknowledgements}
This project has received funding from SES and the Luxembourg National Research Fund (FNR) under the Industrial Partnership Block Grant (IPBG), ref. IPBG19/14016225/INSTRUCT. 
\end{acknowledgements}

\bibliographystyle{spbasic}
\balance

\end{document}